\documentclass[a4paper,fleqn]{cas-sc}

\usepackage[authoryear]{natbib}
\usepackage[utf8]{inputenc}
\usepackage{graphicx}	% Including figure files
\usepackage{amsmath}	% Advanced maths commands
\usepackage{amssymb}	% Extra maths symbols
\usepackage{latexsym}
\usepackage{hyperref}	% Permits to insert hypertext links
\usepackage{mathtools}
\usepackage{graphicx}
\usepackage{aas_macros}
\usepackage{subcaption}
\usepackage{rotating}
\usepackage{epsfig}
\usepackage{tabularx}
\usepackage{natbib}
\usepackage{xcolor}
\usepackage{booktabs}
\usepackage{comment}
\usepackage[T1]{fontenc}
\usepackage[english]{babel}

\def\tsc#1{\csdef{#1}{\textsc{\lowercase{#1}}\xspace}}
\tsc{WGM}
\tsc{QE}
\begin{document}
%\linenumbers
\let\WriteBookmarks\relax
\def\floatpagepagefraction{1}
\def\textpagefraction{.001}

\shorttitle{Pre-impact strewn field predictions}    

\shortauthors{Moscati et al.}  

% \title [mode = title]{From orbit to ground: pre-impact meteorite strewn field predictions for planetary defence}  
\title [mode = title]{From orbit to ground: pre-impact meteorite strewn field predictions for imminent impactors and meteorite recovery}  




\author[1,2]{Anna Moscati}[orcid=0009-0002-2191-2845]

\cormark[1]


\ead{anna.moscati@outlook.com}

%\ead[url]{www.esa.int}

\credit{Conceptualization, Methodology, Software, Formal analysis, Investigation, Writing – original draft, Writing – review and editing, Visualization.}

\affiliation[1]{organization={Technische Universiteit Delft},
            addressline={Mekelweg 5}, 
            city={Delft},
%          citysep={}, % Uncomment if no comma needed between city and postcode
            postcode={2628}, 
%            state={XXX},
            country={The Netherlands}}

\affiliation[2]{organization={ESA ESRIN/PDO/NEO Coordination Centre},
            addressline={Largo Galileo Galilei, 1}, 
            city={Frascati (RM)},
            %citysep={}, % Uncomment if no comma needed between city and postcode
            postcode={00044}, 
            country={Italy}}

\author[2]{Marco Fenucci}[orcid=0000-0002-7058-0413]

% Footnote of the second author

% Email id of the second author
\ead{marco.fenucci@ext.esa.int}

%\ead[url]{https://neo.ssa.esa.int/}

\credit{Conceptualization, Supervision, 
Methodology, Validation, Writing – original draft}

\author[2]{Laura Faggioli}[orcid=0000-0002-5447-432X]

% Footnote of the second author

% Email id of the second author
\ead{laura.faggioli@ext.esa.int}

\credit{Conceptualization, Supervision, 
Methodology, Writing – review and editing}
%\affiliation[2]{organization={},
%          citysep={}, % Uncomment if no comma needed between city and postcode

\author[2]{Marco Micheli}[orcid=0000-0001-7895-8209]

% Footnote of the second author

% Email id of the second author
\ead{marco.micheli@ext.esa.int}

\credit{Validation, Writing – review and editing}
%\affiliation[2]{organization={},
%          citysep={}, % Uncomment if no comma needed between city and postcode

\author[3]{Francisco Ocaña}[orcid=0000-0002-9836-3285]

% Footnote of the second author

% Email id of the second author
\ead{francisco.ocana@ext.esa.int}

\credit{Validation, Writing – review and editing}
\affiliation[3]{organization={ESA ESAC/PDO},
            addressline={Bajo del Castillo s/n}, 
            city={Villafranca del Castillo, Madrid},
          citysep={}, % Uncomment if no comma needed between city and postcode
            postcode={28692}, 
            %state={},
            country={Spain}}

\author[4]{Juan Luis Cano}[orcid=0000-0002-2005-4255]

% Footnote of the second author

% Email id of the second author
\ead{juan-luis.cano@esa.int}

\credit{Validation, Writing – review and editing}
\affiliation[4]{organization={ESA ESOC/PDO},
            addressline={Robert-Bosch-Straße 5}, 
            city={Darmstadt},
          citysep={}, % Uncomment if no comma needed between city and postcode
            postcode={64293}, 
            %state={},
            country={Germany}}

\cortext[1]{Corresponding author}

\fntext[1]{}


\begin{abstract}
The flux of meteoroids reaching the Earth is continuous, ranging from microscopic
grains to occasional metre and decametre scale bodies. The smallest ones fully ablate 
in the upper atmosphere, whereas sufficiently large or strong objects survive entry 
and deposit fragments on the ground as meteorites. Predicting where these fragments 
land, and reconstructing the atmospheric trajectory and fragmentation sequence that 
produced them, is central both to hazard assessment and to the recovery of freshly 
fallen material. The accuracy of such predictions, however, remains limited by 
poorly constrained fragmentation processes and by sparse, heterogeneous observational 
coverage of individual events. Traditional strewn field simulations rely on detailed fireball data and event-specific assumptions on fragment masses, aerodynamics, and breakup. These approaches are effective for well-instrumented events, but their applicability degrades rapidly when observations are sparse, often resulting in huge uncertainties. We present an \emph{ab initio} framework predicting strewn fields of near-Earth asteroids directly from pre-impact orbital solutions. It propagates luminous trajectory and dark flight using a physics-based translational dynamics model and realistic atmospheric conditions, without requiring fireball triangulation or event-specific tuning. Validation against recent asteroid falls with recovered meteorites shows agreement with observations, with nominal solutions reproducing fall locations within 100-200 m. The new method has been integrated into the ESA Aegis pipeline, which now enables hours-ahead computation of impact locations, supporting recovery efforts, minimizing contamination, and, where warranted by object size and predicted ground hazard, civil-protection decision making.
\end{abstract}


\begin{highlights}
\item We developed a Monte Carlo model for the computation of the strewn field of an impacting meteoroid
\item We integrated the model into the pipeline for imminent impactors monitoring of the European Space Agency
\item Tests performed on previous impactors indicate that the model is reliable for meteorite recovery purposes 
\end{highlights}

% Keywords

\maketitle

\section{Introduction}
\label{s:intro}

  The Earth is constantly bombarded by meteoroids of varying sizes and compositions. Most disintegrate at high altitudes, but a small fraction survives atmospheric ablation and reaches the surface; approximately two $10^{3}$~kg impacts occur every year, with a $10^{5}$~kg fall every $\sim$100 years \citep{2006M&PS...41..607B}. 
Events such as the Carancas impact \citep{2008A&A...485L...1B} and the Chelyabinsk 
airburst \citep{popova2013} demonstrated that even meter to decametre scale objects can produce observable ground consequences. In particular, the Carancas impact 
produced a crater and triggered toxic gas emissions that affected the local population, while the Chelyabinsk event caused more than a 
thousand people injured,  mainly through secondary effects of the shock wave such as shattered glass.
The lack of specific predictions prior to atmospheric entry limited the ability to anticipate the affected area and to coordinate response measures. For metre-scale imminent impactors, the expected consequences are typically small and local, but prior knowledge of the possible impact region and of the properties of surviving fragments could still improve situational awareness and response times for meteorite-recovery teams, civil-protection authorities, and planetary defence services.
%

% Introduction to imminent impactors
  Over the past decade, the capability to identify asteroids prior to impact has improved dramatically, due to the development of dedicated surveys for near-Earth asteroids (NEAs). NASA has led this effort with the Catalina Sky Survey (CSS), the 
Panoramic Survey Telescope and Rapid Response System (Pan-STARRS), and the Asteroid 
Terrestrial-impact Last Alert System (ATLAS), while the European Space Agency (ESA) 
is developing the upcoming Flyeye telescope \citep{arcidiacono2024}. The Vera Rubin Observatory is expected to further strengthen this capability during its ten-year operational lifetime \citep{frazer-etal_2026}. At the same time, individual observers have an important role: for example, Krisztián Sárneczky discovered the small impactor 2023~CX1 prior to atmospheric entry, enabling rapid follow-up observations and orbit determination \citep{2026AcAau.240....1S}.
On the other hand, monitoring systems such as ESA’s Aegis/Meerkat\footnote{\url{https://neo.ssa.esa.int/}}, NASA’s Sentry/Scout\footnote{\url{https://cneos.jpl.nasa.gov/}}, and NEODyS/NEOScan\footnote{\url{https://newton.spacedys.com/neodys/}}, originally developed at the University of Pisa and SpaceDyS s.r.l.\footnote{\url{https://www.spacedys.com/}}, now allow impact probabilities to be evaluated almost in real time. Thanks to this infrastructure, 11 NEAs have been identified prior to atmospheric entry as of February 2026, with a median warning time of about 9~h. Owing to this very short notice, these objects are commonly referred to as \emph{imminent impactors}. For such cases, predicting the strewn field before entry represents the final step in the planetary defence chain. While fall models were tested \citep{2025Icar..42516345C}, and a real time strewn field prediction attempt was performed for the case of the 2024~XA1 impactor \citep{gianotto-etal_2025}, a fully automated system is not in place yet.

% Scientific interest of meteorites
  Meteorite falls are of particular scientific interest because recovered samples provide direct material from small Solar System bodies. When the pre-impact heliocentric orbit is known, detailed laboratory analyses can constrain their origin and the properties of their parent bodies \citep{2010M&PS...45.1557S, 2022M&PS...57.1328D, 2024A&A...686A..67S, 2025NatAs...9.1624E}, improving our understanding of Solar System formation and evolution. Predicting the strewn field in advance would further enable rapid recovery before significant terrestrial alteration.
% Modelling the entry in the atmosphere: phases and challenges
  Modelling the atmospheric entry and ground dispersion of asteroids is a challenging problem due to the intrinsically complex and coupled physical processes involved \citep{2021MNRAS.503.3337M}. As a meteoroid enters the atmosphere at hypersonic velocity, its trajectory and mass evolution are governed by aerodynamic drag, thermal ablation, and mechanical fragmentation, all occurring within a rapidly changing atmospheric environment. Fragmentation typically takes place when the dynamic pressure exceeds the material strength of the body \citep{Chyba1993}, a parameter that is poorly constrained and can vary by orders of magnitude depending on the internal structure and composition of the object.
  After breakup, the surviving fragments experience a transition from the luminous trajectory to the dark flight phase, during which their motion becomes highly sensitive to atmospheric density and wind fields. Small variations in fragmentation altitude, fragment mass, or wind conditions can lead to large differences in the final impact locations, often amounting to hundreds of metres or several kilometres on the ground. As a consequence, the prediction of strewn fields is affected by significant uncertainties and cannot be treated as a purely deterministic problem, especially in the absence of detailed observational constraints.

    Previous studies have addressed related aspects of atmospheric fragmentation, strewn-field reconstruction, crater-forming fragments, and impact-hazard modelling. For example, \cite{Bronikowska2017} reconstructed the Morasko crater strewn field, \cite{Schmalen2022} modelled the atmospheric entry of the Campo del Cielo iron meteoroid, and \cite{Luther2023} further investigated the formation of Campo del Cielo funnels and craters. More recently, \cite{Luther2026} discussed impact hazards from small iron NEAs, while \cite{Wheeler2024} addressed probabilistic impact-hazard modelling over a broader size range. These works provide important context for the present study, but they primarily concern post-event reconstruction, crater/funnel formation, or damage assessment rather than pre-impact meteorite strewn-field prediction from orbital information.

% Strewn fields from all-sky cameras
  The vast majority of successful strewn field reconstructions rely on observations of atmospheric fireballs, typically provided by networks of all-sky cameras such as the Prima Rete Italiana per la Sorveglianza sistematica di Meteore e Atmosfera (PRISMA; \citealt{2016pimo.conf...76G}), the Fireball Recovery and InterPlanetary Observation Network (FRIPON; \citealt{2020A&A...644A..53C}), the European Fireball Network (\citealt{1998M&PS...33...49O,2022A&A...667A.158B}), the Desert Fireball Network (DFN; \citealt{2017ExA....43..237H,2019MNRAS.483.5166D}), the Spanish Meteor Network (SPMN; \citealt{2001JIMO...29..139T}), and many others. Classical formulations, originating from the work of \citet{1987BAICz..38..222C}, combine triangulated fireball trajectories, deceleration measurements, and photometric data to estimate the terminal state of the meteoroid, which then provides the initial conditions for dark-flight simulations. 
Numerous recent studies rely on this fireball-derived terminal state as input and instead focus on determining the strewn field through different modelling approaches \citep[e.g.][]{Bronikowska2017, 2021MNRAS.503.3337M, 2021Icar..36714553L, 2022M&PS...57.1328D, 2022M&PS...57.2108S, 2022PSJ.....3...44T, 2024A&A...686A..67S}.
While highly effective, this approach is inherently event-specific and critically dependent on the availability and quality of observations, and it is not directly applicable when only pre-impact orbital information is available. In addition, the full reconstruction pipeline, from data acquisition to trajectory fitting and model calibration, must be completed before any actionable search area can be defined, thus introducing time lags between the fall and the meteorite searching campaign. Moreover, many of these implementations rely on simplifying assumptions, such as a single dominant fragmentation event without cascade breakup, neglect of atmospheric winds and cross-range transport, and fragment masses that are prescribed rather than self-consistently derived.

  The limitations outlined above motivate the development of an \emph{ab initio} approach such as that by \citet{2025Icar..42516345C}. Building on this preliminary study, here we introduce a fully numerical, end‑to‑end framework that predicts meteorite strewn fields directly from orbital solutions, without requiring any fireball data. This allows the computation of predicted strewn fields even before impact, when such orbital data is available. Starting from a heliocentric orbit, we reconstruct the high‑altitude entry state and propagate the body through both the luminous and dark‑flight phases.  
Fragmentation is treated as a cascading, statistically defined process that can undergo multiple generations of break‑up, with lateral ejection velocities assigned to child fragments.
Real-time and forecasts for the global atmospheric conditions, which are needed for the case of imminent impactors, are retrieved from the Global Forecast System (GFS). Section~\ref{s:model} covers these aspects of the entry model. 
Dynamics and fragmentation are gathered together in a Monte Carlo framework to model uncertainties in entry state, material properties, fragmentation parameters, and atmospheric conditions, to obtain statistically robust strewn field probability maps (Sec.~\ref{s:montecarlo}). The Monte Carlo method is implemented within an end-to-end pipeline linking ESA Meerkat identification of imminent impactors, with ESA Aegis orbit propagation and strewn field probability maps using the present solver (Sec.~\ref{s:meerkataegis}). The model is validated against several well‑observed meteorite falls with recovered strewn fields, demonstrating sub‑kilometre agreement between simulated impacts and recovered meteorites (Sec.~\ref{s:results}). 
We then discuss limitations and possible improvements of the model in Sec.~\ref{s:discussion}, and give the conclusions in Sec.~\ref{s:conclusions}.

\section{Fall model and numerical framework}
\label{s:model}
\subsection{Overview of the entry–fragmentation–fall process}
  When a small near-Earth asteroid\footnote{In the rest of the paper, we will use both terminologies \emph{small near-Earth asteroid} or \emph{meteoroid} with the same meaning. This ambiguity is consistent with the definition of meteoroid given by the International Astronomical Union \citep{IAU2024} and the definition of near-Earth asteroid given by the Minor Planet Center.} (NEA) intersects the terrestrial atmosphere, its evolution is governed by a coupled sequence of aerodynamic deceleration, thermal mass loss and fragmentation, ultimately leading to either complete ablation or the formation of a strewn field.

 At the onset of atmospheric entry, the body encounters increasingly dense air and 
begins to heat up while still travelling at hypervelocities. For asteroidal material, 
geocentric entry speeds are bounded below by Earth's escape velocity 
($\sim 11.2\,\mathrm{km\ s^{-1}}$), with a practical upper limit of 
$\sim 30\,\mathrm{km\ s^{-1}}$ for meteorite-dropping events 
\citep{Ceplecha1998}; for instance, the 13 recovered meteorite-dropping 
falls compiled by \citet{2011M&PS...46.1525P} span entry velocities of 
$12.4\text{--}22.5\,\mathrm{km\ s^{-1}}$. Strong aerodynamic drag and the formation of a bow shock lead to rapid compression and heating of the surrounding gas, initiating ablation and producing the luminous fireball phase. During this regime, the trajectory and mass evolution are primarily controlled by aerodynamic loading and heat transfer, and the meteoroid retains much of its pre-atmospheric velocity \citep{Duffa2017MeteorsEntry}.

  As atmospheric density increases, the dynamic pressure acting on the body may exceed its mechanical strength, triggering fragmentation \citep{Chyba1993}. Depending on the type of material, this process can range from the separation into a small number of large fragments (like for iron meteoroid) to the production of a large population of smaller pieces and debris (like for stony and carbonaceous chondrite) \citep{2020AJ....160...42B}. Fragmentation substantially enlarges the effective cross-sectional area interacting with the atmosphere, increasing deceleration, ablation, and energy dissipation. In strongly disruptive cases, this rapid energy loss may manifest as an airburst, whereas more modest event leads to a progressive decoupling of individual fragments, which develop their shock wave as single fireballs that continue their entry independently.

  Following sufficient deceleration, typically once velocities fall below a few kilometres per second \citep{2021MNRAS.503.3337M}, luminous emission ceases and the motion transitions into the dark-flight regime. In this phase, the fragments have lost all memory of their initial velocity and behave as non-luminous bodies falling through the lower atmosphere. The interaction with three-dimensional wind fields becomes the dominant factor controlling lateral dispersion and ultimately determines the spatial distribution of impact points on the ground.

\subsection{Dynamical modelling}
 
The three-dimensional continuous numerical propagation of the asteroid trajectory is performed in an Earth-centered inertial (ECI) reference frame using Cowell’s method with an adaptive Runge–Kutta–Fehlberg 4(5) integrator, implemented in the \texttt{Tudat} software\footnote{\url{https://docs.tudat.space}} \citep{2025epsc.conf..673D}. Initial mass, position and velocity vectors at 100 km from the surface are used as starting conditions for the propagation. The Earth is modeled as a WGS84 ellipsoid, and only central gravitational attraction and aerodynamic drag are included in the equations of motion; a preliminary analysis showed that the effects of higher-order spherical harmonics and third-body perturbations are negligible for the considered cases. Given the extreme variability in meteoroid shapes and the highly irregular geometries of the resulting fragments, a detailed shape-resolved aerodynamic treatment is neither feasible nor observationally constrained. Therefore, both the parent body and its fragments are assumed to be spherical. Under this assumption, lift and side forces vanish by symmetry, and the atmospheric entry can be treated as purely ballistic, with aerodynamic drag being the only force acting on the body.

  The propagation is modeled using translational motion only, while rotational dynamics are neglected. These effects are generally considered secondary and become relevant only if significant spin or tumbling develops, introducing additional centrifugal effects and substantially increasing model complexity \citep{1987BAICz..38..222C}. 
Although fragment rotation has been discussed as possible 
source of lateral dispersion \citep{1980Icar...42..211P} and modelled numerically 
by \citet{Azovskii2002}, who found them to contribute to the debris scatter 
at least as much as the aerodynamic interaction between fragments, such effects are 
rarely incorporated into operational strewn field prediction models 
\citep{1987BAICz..38..222C}. Because aerodynamic drag and wind dominate during dark 
flight and largely determine strewn field accuracy, secondary effects of this kind, 
including Magnus-type forces, are typically neglected, as their influence is 
relatively small and unlikely to meaningfully improve results.

  The system is formulated as a non-linear differential equation in state-vector form,
\begin{equation}
\mathbf{y} =
\begin{bmatrix}
\mathbf{r} \\
\mathbf{v} \\
m
\end{bmatrix},
\qquad
\dot{\mathbf{y}} =
\begin{bmatrix}
\mathbf{v} \\
- \mu \dfrac{\mathbf{r}}{|\mathbf{r}|^3}
- \dfrac{1}{2m} \rho_{\rm air} C_d(M) A_0 |\mathbf{V}_r| \mathbf{V}_r \\
- \dfrac{1}{2} \sigma_{\rm abl} \rho_{\rm air} A_0 |\mathbf{V}_r|^3 \;\; \text{if } |\mathbf{V}_r| \ge 3~\mathrm{km\,s^{-1}}, \;\; 0 \;\; \text{otherwise}
\end{bmatrix},
\end{equation}
where $\mathbf{r}$ and $\mathbf{v}$ are the position and velocity vectors in the Earth-Centered Inertial (ECI) frame, $m$ is the instantaneous mass, $\mu = G M_\oplus = 3.986 \times 10^{14}\ 
\mathrm{m^3\,s^{-2}}$ is the Earth gravitational parameter, $G$ the gravitational 
constant, $M_\oplus$ Earth's mass, $\rho_{\rm air}$ is the atmospheric density, $C_d(M)$ the drag coefficient as a function of Mach, $A_0$ the reference projected area, $\sigma_{\rm abl}$ the ablation coefficient, and $\mathbf{V}_r=\mathbf{v}-\boldsymbol{\omega}_\oplus\times\mathbf{r}$ the velocity vector relative to the co-rotating atmosphere, with $\boldsymbol{\omega}_\oplus$ the instantaneous Earth's rotation vector. The Earth orientation and the transformation between the Earth-fixed and ECI frames are computed using the SPICE reference-frame and Earth-orientation models. Here $M = |\mathbf{V}_r|/a$ is the Mach number, where 
$a = \sqrt{\gamma R_{\rm s} T}$ is the local speed of sound, $\gamma = 1.4$ is the 
heat capacity ratio of air, $R_{\rm s} = 287\ \mathrm{J\,kg^{-1}\,K^{-1}}$ is the 
specific gas constant for dry air, and $T$ is the local atmospheric temperature.

\begin{figure}
    \centering
    \includegraphics[width=0.6\linewidth]{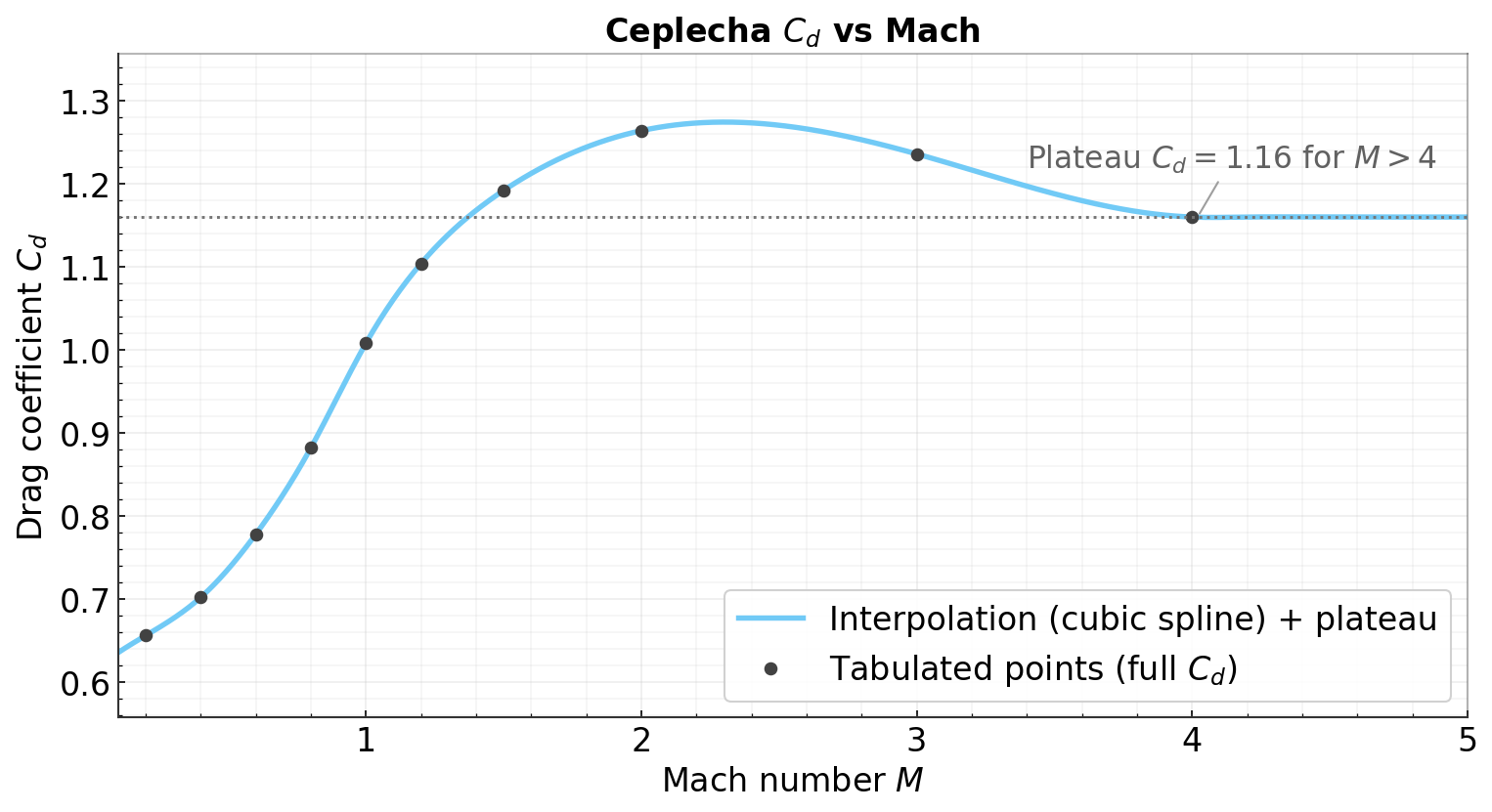}
    \caption{Mach-dependent drag coefficient $C_d$ following the Ceplecha formulation. Tabulated values are interpolated using a cubic spline, with a constant plateau $C_d = 1.16$ assumed for $M > 4$.}
    \label{fig:cdvsmach}
\end{figure}

  The acceleration consists of two contributions: gravity and aerodynamic drag. The latter is active throughout the entire trajectory and is modeled using a Mach-dependent coefficient following \citet{1987BAICz..38..222C}, obtained via cubic spline interpolation between tabulated points, as shown in Fig.~\ref{fig:cdvsmach}. 
  Ablation is treated as a continuous mass loss occurring exclusively during the fireball phase and is described by the scalar equation in \cite{Chyba1993}. The process is assumed to stop at the onset of dark flight, which is identified when the velocity relative to the atmosphere drops below $3~\mathrm{km\ s}^{-1}$ \citep{2021MNRAS.503.3337M}. This threshold was adopted as a commonly used operational criterion in meteorite strewn-field and dark-flight modelling. It is not meant to imply a sharp physical discontinuity in the flight regime, but rather to define the velocity range in which luminous ablation is expected to become negligible and the subsequent trajectory is governed primarily by aerodynamic drag, gravity, and atmospheric winds. The coefficient $\sigma_{\rm abl}$ encapsulates the combined effects of aerodynamic heating and material response and can be expressed as the ratio between the heat transfer coefficient $C_H$ and the effective heat of ablation $Q$.
The parameter $Q$ represents the energy required to remove unit mass from the body and accounts for both fusion and vaporisation processes \citep{1980Icar...42..211P}. For stony meteoroids, commonly adopted values are $Q \simeq 8 $~MJ~kg$^{-1}$ and $C_H \simeq 0.1$, consistent with classical meteor ablation models in \cite{Chyba1993, Avramenko2014, Johnston2018}. However, both quantities are subject to significant uncertainty. Experimental and observational studies suggest that lower values of $Q$, of the order of $4$~MJ~kg$^{-1}$, may be more appropriate under certain conditions \citep{Stern2017}, while the heat transfer coefficient has been shown to decrease at altitudes below approximately 30~km and may reach values as low as $C_H \sim 0.05$ \citep{Johnston2018}.
As a consequence, the ablation coefficient $\sigma_{\rm abl}$ cannot be considered a well-constrained physical constant and is expected to vary depending on the meteoroid composition, structure, and flow regime. In the present work, this uncertainty is explicitly accounted for by treating $\sigma_{\rm abl}$ as a stochastic parameter sampled from a specified distribution within a Monte Carlo framework (as in Table \ref{tab:physical_parameters}), rather than adopting a single nominal value.

  The propagation is terminated upon dynamic-pressure-induced fragmentation, ground impact on the ellipsoid, or complete mass depletion ($m \le 1$~g). Fragmentation events generate a cascade of child fragments that are processed sequentially in a queue-based framework, with each fragment propagated independently until termination. For computational efficiency, fragments below a certain mass threshold are grouped into log-spaced bins, with each bin represented by a single particle whose mass and multiplicity preserve the statistical distribution of the original fragments.

  Since the numerical integrator nominally detects ground impacts when the altitude relative to the WGS84 reference ellipsoid approaches zero, a post-processing correction is applied to accurately determine the true terrain intersection. For each trajectory, a terrain-aware impact refinement is performed by querying the ALOS World 3D 30m digital elevation model (DEM) through the OpenTopography interface made by the \cite{AW3D30_OpenTopo}. Local surface elevation is obtained via bilinear interpolation on the DEM grid using geodetic latitude and longitude. To achieve sub-step accuracy, the propagated state vector and corresponding geodetic coordinates are reconstructed using one-dimensional eighth-order Lagrange interpolation in time. The impact point is then identified as the first sign change of the quantity $(h - h_{\rm DEM})$ along the trajectory, and the exact crossing time and location are recovered by linear interpolation.

\subsection{Fragmentation modelling}
  During the luminous phase, meteoroid fragmentation is governed by a ram pressure criterion, in which disruption occurs when the dynamic term exceeds the material strength of the body \citep{Chyba1993}, i.e. when
\begin{equation}
p_{\rm ram} \simeq \frac{1}{2} \rho_{\rm air} |\mathbf{V}_r|^2 \ge S,
\end{equation}
where $S$ is the internal strength of the meteoroid. At hypervelocity, static pressure is negligible and the ram pressure is dominated by the dynamic term. The value of $S$ therefore controls the altitude of the initial breakup, and successive fragmentation events may follow, allowing the progressive disruption of the meteoroid to be represented during the luminous flight.

\paragraph{Mass distribution}
A fragmentation event is modelled in the following way: a small fraction of mass, referred to as \emph{dust} (untracked debris with mass $\le 1$~g), is removed from consideration. Let $M_{\mathrm{children}}$ denote the remaining mass available for discrete fragments. The fragment distribution follows a cumulative power-law formulation commonly adopted for fragmentation processes \citep{Bronikowska2017} and is implemented here following the analytical formulation discussed by \citet{2024P&SS..24105838B}. The normalized masses are defined as $\mu_k=m_k/M_{\mathrm{children}}$ and $\mu_l=m_l/M_{\mathrm{children}}$, where $k$ denotes the cumulative fragment rank in decreasing order of mass. The parameter $n_l$ specifies the number of co-largest fragments and can be set as an input parameter, with $n_l=1$ adopted as the default value. Their common maximum mass is defined as $m_l$. The first $n_l$ fragments are therefore assigned the mass $m_l$, while the remaining masses are obtained by inverting the cumulative rank relation,
\begin{equation}
\mu_k=
\left[
\mu_l^{-\beta}
+
\frac{\beta}{1-\beta},
\mu_l^{1-\beta},
(k-n_l)
\right]^{-1/\beta},
\end{equation}
for $k=n_l+1,\ldots,k_{\max}$ and $0<\beta<1$. The dimensional masses are recovered as $m_k=\mu_kM_{\mathrm{children}}$. The total number of fragments is not prescribed \emph{a priori}, but emerges as $k$ is progressively increased until either $m_k < m_{\min} = 1~\mathrm{g}$ or the available mass is exhausted. If the next predicted mass exceeds the remaining available mass, it is set equal to the residual, thereby enforcing $\sum_{k=1}^{k_{\max}}m_k=M_{\mathrm{children}}$.

\begin{figure}
    \centering
\includegraphics[width=0.6\linewidth]{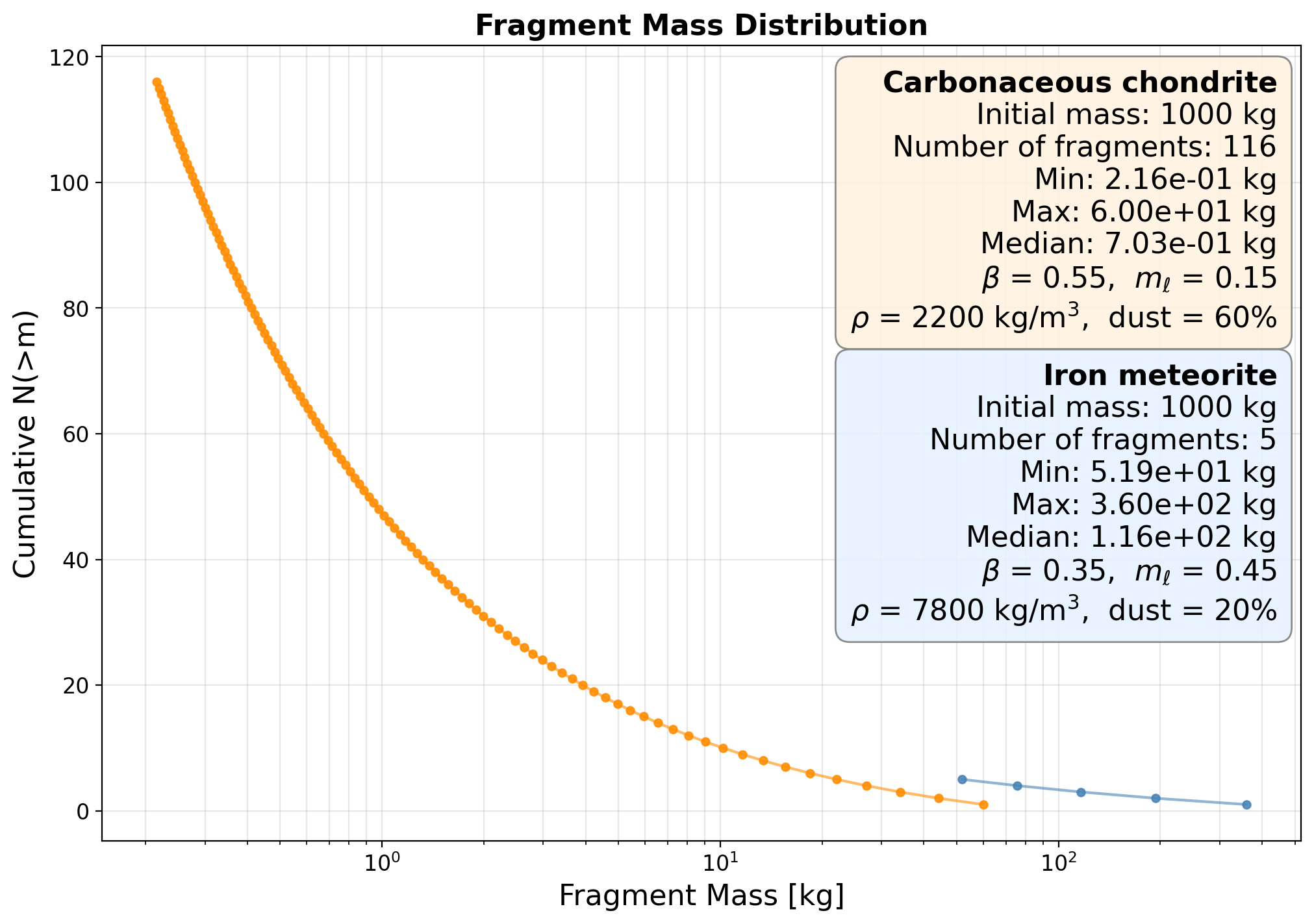}
    \caption{Comparison of cumulative fragment mass distributions for carbonaceous chondrite and iron meteoroid cases with identical initial mass. The curves show the number of fragments with mass greater than m, highlighting the production of many small fragments in the weaker carbonaceous case and fewer, larger fragments in the stronger iron case.}
    \label{fig:mass}
\end{figure}

  The distribution contains several free parameters, allowing the model to adapt to meteoroids of different compositions. For instance, an iron meteoroid typically has a low dust fraction, small $\beta$, and large $m_l$, generating few big fragments or none at all, whereas a carbonaceous chondrite typically has a higher dust fraction, larger $\beta$, and smaller $m_l$, producing numerous small fragments (see Fig.~\ref{fig:mass}). Typically, the meteoroid type is not known prior to impact, therefore the fragmentation parameters are sampled from plausible distributions in a Monte Carlo framework. On the other hand, compositional information may be available after impact through other types of observations. This information can be used after impact to possibly refine the strewn field in order to ease the search for possible meteorite fragments.

\paragraph{Strength distribution}
The strength of the fragments may also not be preserved. Meteoroids are generally characterized by strong internal heterogeneity, with outer layers often consisting of fractured, porous, rubble-pile material and a more compact, coherent interior. During atmospheric entry, mechanical loading induced by aerodynamic pressure preferentially disrupts the weakest regions, causing the failure of low-strength outer layers at higher altitudes. As fragmentation progresses, these heterogeneous components are progressively removed, and the surviving fragments are increasingly representative of stronger, flaw-poor material. As a result, smaller fragments are statistically stronger than their parent bodies \citep{Svetsov1995, Scheeres2015}. This behaviour is consistent with Weibull-type flaw scaling \citep{Weibull1951}, in which the effective material strength increases as the characteristic volume or mass decreases, due to the reduced probability of hosting critical defects. In our fragmentation model, the strength of a child fragment is therefore scaled with respect to that of its parent according to
\begin{equation}
S_{\rm child} = \min \left[ S_{\rm parent} \left( \frac{M_{\rm parent}}{m_{\rm child}} \right)^{\alpha}, \, S_{\rm cap} \right],
\end{equation}
where $M_{\rm parent}$ and $m_{\rm child}$ are the masses of the parent body and the fragment, respectively, $\alpha$ is the Weibull scaling exponent, and $S_{\rm cap}$ represents an upper limit to the achievable strength.

\paragraph{Velocity distribution}
At each fragmentation event, all child fragments inherit the same position and the same velocity vector of the parent body at the breakup point. In addition, an extra velocity component is imparted to each fragment in order to model the lateral dispersion of the strewn field, as shown in Fig. \ref{fig:fragmentation}. Following \cite{1980Icar...42..211P}, we modified their formulation to better reproduce the dispersion of smaller fragments by squaring the ratio of the diameters:
\begin{equation}
|\Delta \mathbf{V}| = \sqrt{\frac{3}{2}\,\frac{\rho_{\rm air}}{\rho_m}\left(\frac{D_{\rm parent}}{d_{\rm child}}\right)^2}\,|\mathbf{V}_r|,
\end{equation}
where $\rho_{\rm air}$ is the atmospheric density at the fragmentation altitude, $\rho_m$ is the bulk density of the meteoroid, which is assumed constant throughout the propagation, and $D_{\rm parent}$ and $d_{\rm child}$ are the diameters of the parent body and the fragment, respectively. The formulation adopted here is based on the lateral-dispersion mechanism of 
\citet{1980Icar...42..211P}, but adapted to a multi-fragment breakup rather than 
to the two-fragment case considered in the original work. 
Their formulation was derived for the separation of two fragments and 
was primarily constrained using the cross-range extent of observed crater fields. 
Its direct application to a population of many fragments spanning a broad range 
of sizes is therefore not expected to provide a unique description of the 
size-dependent lateral dispersion. We consequently retain the physical scaling 
of their formulation while empirically modifying the diameter dependence for 
the multi-fragment regime considered here.
In their formulation, the coefficient $C$ is poorly constrained and was estimated 
from observed crater-field cross-range dispersions to be of order unity, probably 
within a factor of two; we therefore set $C=1$. 
Specifically, the diameter ratio is squared to produce the stronger 
size-dependent dispersion required by the validation cases considered in this 
work. This choice was guided by the validation cases considered in this work, 
for which the resulting terminal velocity distribution and strewn-field morphology 
provided the best agreement with the recovered meteorites. The lateral velocity perturbation is defined relative to the parent fragment velocity at the breakup point. Each child fragment initially inherits the parent position and velocity, and then receives a transverse velocity perturbation. While the magnitude of the additional velocity follows the above expression, its direction is randomly sampled within the plane perpendicular to the parent velocity vector, as can be seen in Fig.~\ref{fig:fragmentation}. The perturbation is then added to the parent velocity in the inertial frame in which the trajectory is propagated. Since the perturbation direction is sampled in the plane perpendicular to the parent velocity vector, the additional velocity represents lateral spreading of the fragment cloud rather than a change in the along-track velocity by construction. As a consequence, smaller fragments acquire larger transverse velocity components, leading to an increasing lateral spread of the fragment cloud. {Without this transverse velocity dispersion, fragments of different masses would remain confined to an approximately deterministic downrange curve, as in the formulation of \cite{2025Icar..42516345C}. The lateral perturbation therefore introduces the cross-range dispersion required to reproduce the morphology of observed strewn fields.
Because multiple fragments are generated at breakup, conservation of linear momentum is enforced explicitly. After the individual perturbations $\delta \mathbf{v}_i$ are generated, their mass-weighted mean is removed from each fragment,
\begin{equation}
\delta \mathbf{v}_i \leftarrow \delta \mathbf{v}_i -
\frac{\sum_j m_j \delta \mathbf{v}_j}{\sum_j m_j},
\end{equation}
which ensures
\begin{equation}
\sum_i m_i \delta \mathbf{v}_i = 0
\end{equation}
by construction. As a result, the centre-of-mass velocity of the resolved fragment cloud remains equal to the velocity of the parent body immediately before breakup. The unresolved dust component is assumed to continue with the parent velocity, so that the total linear momentum of the resolved fragments plus dust is conserved at the fragmentation event. This correction removes the spurious net lateral velocity of the fragment cloud while preserving the relative velocity differences between fragments, and therefore the physical lateral dispersion.

\begin{figure}
    \centering
    \begin{subfigure}{0.48\linewidth}
        \centering
        \includegraphics[width=\linewidth]{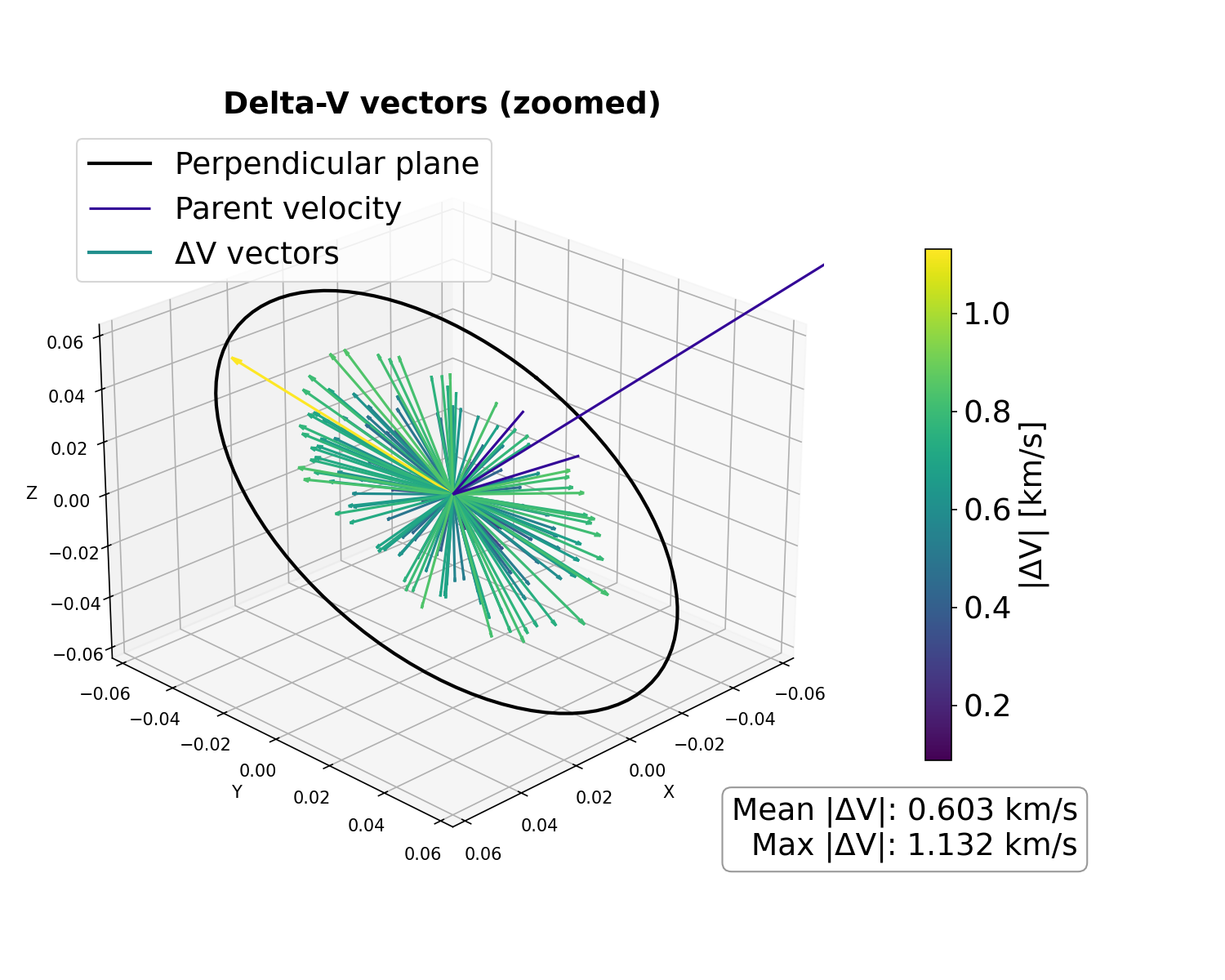}
        \label{fig:frame3}
    \end{subfigure}
    \hfill
    \begin{subfigure}{0.48\linewidth}
        \centering
        \includegraphics[width=\linewidth]{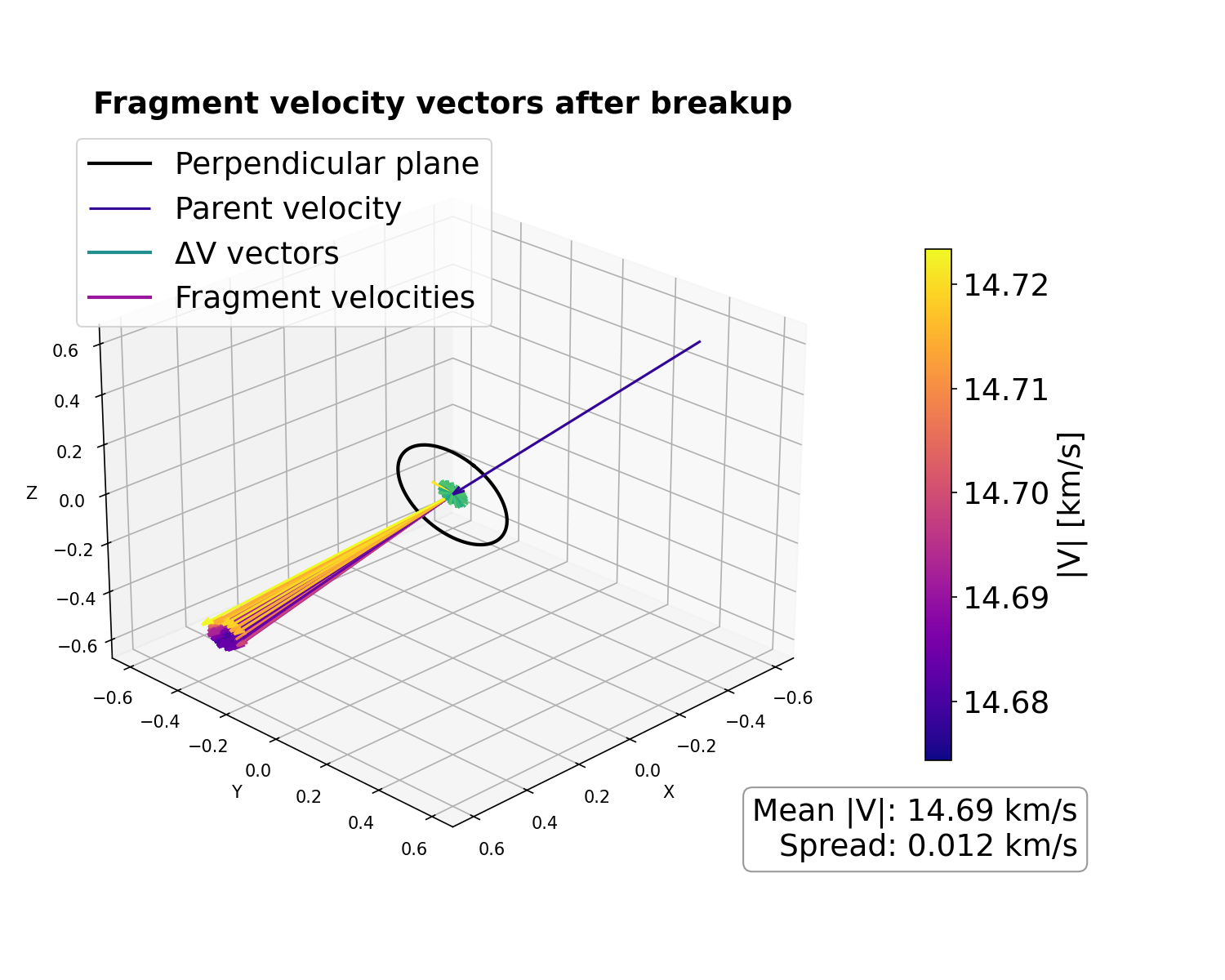}
        \label{fig:lastframe}
    \end{subfigure}
    \caption{Example of fragmentation kinematics. Left: distribution of post-breakup $\Delta V$ vectors in the plane perpendicular to the parent velocity, colored by magnitude. Right: resulting fragment velocity vectors obtained by adding $\Delta V$ to the parent velocity, colored by total speed; the black circle indicates the perpendicular plane and the blue arrow the parent trajectory.}
    \label{fig:fragmentation}
\end{figure}

  During atmospheric flight, the aerodynamic deceleration of each fragment is governed by its area-to-mass ratio, which for spherical bodies scales as $A_0/M \propto 1/R$, implying stronger drag for smaller bodies. The additional lateral velocity imparted at fragmentation further increases their total velocity and, consequently, their deceleration, causing smaller fragments to fall earlier along the trajectory, while larger ones propagate farther downrange. This combined effect produces an elongated strewn field, with larger bodies aligned along the trajectory and smaller ones deposited closer to the entry point with a broader lateral dispersion, in the absence of significant aerodynamic interactions or complex flow effects, such as those observed in the case of 2008~TC$_3$ \citep{Jenniskens2022}.

\subsubsection{Fragmentation modelling of decameter sized impactors}
  The above fragmentation implementation is suitable for asteroids with a diameter up to a size of $\sim$10 m. 
For decametric bodies, such as the Tunguska impactor, other processes need to be taken into account to study atmospheric entry. One example is the classical pancake model \citet{Chyba1993, Hills1993}, originally developed over thirty years ago. In its classical formulation, a large body fragments 
in such a way that a coherent shock front spreads across the fragments, which are 
treated as a single deformable cloud that expands laterally under aerodynamic 
pressure. In this pure form the model yields a single airburst/terminal point rather 
than a discrete strewn field. To obtain a strewn field that could be tested against 
observations, we adopted a modified version in which the lateral expansion is 
followed only until the cloud expansion ratio exceeds a prescribed threshold (the 
pancake factor); beyond this point the fragments are assumed to decouple and are 
propagated individually through dark flight. We applied this modified model to the 
Chelyabinsk fall and found that, despite its added complexity, it provides no 
significant improvement in reproducing the strewn field over our simpler 
fragment-based propagation.

  For objects of this size we instead adopted an alternative breakup approach, which extends the methodology described for smaller asteroid. Upon fragmentation, the body is separated into: 1) a core retaining 85\% of the original mass, 2) a dust fraction and, 3) the remaining material, which is treated according to the standard breakup generation rules explained above. The core is assumed to resist further fragmentation (continuing to experience deceleration and mass loss), while the rest of the mass follows the same procedure already described. The adopted 85\% core mass fraction was calibrated using the Chelyabinsk event. Note that this is the only case available for validation; therefore, its applicability could not be independently assessed and may require adjustment or a Monte Carlo treatment in the case of future events, for which the software is already capable to model.

\subsection{Atmospheric modelling}

  The atmospheric model is based on GFS \citep{GFS_NOAA} tabulated profiles with a horizontal resolution of $0.25^\circ$. GFS is a free and publicly available global dataset providing atmospheric forecasts at approximately 3-hour intervals, making it suitable for modeling imminent impactors and estimating strewn fields prior to atmospheric entry. For each simulation, the GFS file closest in UTC time to the target epoch is automatically selected. The dataset provides altitude-dependent profiles of density, pressure, temperature, and horizontal wind components ($u$ and $v$) up to altitudes of approximately 40–50 km. Since the fall model starts at 100 km, properties above the maximum altitude covered by GFS are modeled using the 1976 U.S. Standard Atmosphere \citep{US76}. In this upper region, wind velocities are assumed to gradually decrease to zero from the highest available GFS level. This assumption is justified by the fact that the strongest atmospheric winds occur in the upper troposphere, where jet streams associated with the polar vortex can reach velocities of up to $\sim$100 m s$^{-1}$. At higher altitudes, including the stratosphere and mesosphere, wind velocities are significantly smaller and their influence on the trajectory is negligible compared to the hypersonic entry speed of the incoming body \citep{DOWLING2007169}. A full three-dimensional wind field is reconstructed by bilinear interpolation of the horizontal wind components on the latitude–longitude grid and cubic spline interpolation in the vertical direction. The resulting interpolators are evaluated at each propagation step throughout the simulation.

\section{Monte Carlo Framework}
\label{s:montecarlo}
  The strewn field prediction is based on a Monte Carlo framework that propagates uncertainties in the entry state, atmospheric conditions, and physical properties of the impactor. The parameters are described by a priori distributions representative of different meteoroid types, including ordinary chondrites, carbonaceous, and iron bodies, as summarized in Table~\ref{tab:physical_parameters}.

  The Monte Carlo simulations are organized in batches of $n$ runs. Each batch produces a two-dimensional probability map of the predicted strewn field, obtained by aggregating the individual impact locations and terminal masses. Convergence of the solution is assessed by measuring the similarity between successive probability maps using the Jensen-Shannon divergence \citep{Lin1991JSD}. In practice, the total number of Monte Carlo runs cannot exceed 1000, with convergence typically achieved well before this upper limit. The simulation is considered converged when the divergence falls below a prescribed threshold $\varepsilon$ for $N$ consecutive batches, both in terms of impact location and mass distribution. This criterion ensures that the resulting strewn field map is statistically stable and insensitive to further sampling.

\begin{table}[h]
\centering
\begin{tabularx}{\textwidth}{p{2.7cm} p{3.5cm} l l l}
\toprule
\textbf{Parameter} & \textbf{Distribution} & \textbf{Range (OC)} & \textbf{Range (CC)} & \textbf{Range (Iron)} \\
\midrule
\multicolumn{5}{l}{\textit{Aegis input parameters}} \\
Position $r$ & Multivariate normal from Aegis & --- & --- & --- \\
Velocity $v$ & Multivariate normal from Aegis & --- & --- & --- \\
\midrule
\multicolumn{5}{l}{\textit{Meteoroid physical properties}} \\
Density $\rho$ & Uniform & 2500--3500 kg\,m$^{-3}$ & 2100--3100 kg\,m$^{-3}$ & 7000--7800 kg\,m$^{-3}$ \\
Albedo $p$ & Triangular & 0.075 / 0.208 / 0.33 & 0.02 / 0.06 / 0.18 & 0.07 / 0.18 / 0.54 \\
Diameter $D$ & Derived & --- & --- & --- \\
Mass $m_0$ & Derived & --- & --- & ---\\
Strength $S$ & OC: Bimodal uniform; others: Uniform 
& $[0.04,0.12] \cup [0.9,5]\ \mathrm{MPa}$ 
& 0.25--0.7 $\mathrm{MPa}$ 
& 5--50 $\mathrm{MPa}$ \\
\midrule
\multicolumn{5}{l}{\textit{Fragmentation parameters}} \\
Dust fraction & Uniform & 0--0.9 & 0.3--0.95 & 0.0--0.25 \\
Largest fragment frac. $m_l$ & Uniform & 0.05--0.4 & 0.01--0.25 & 0.4--0.9 \\
$\beta$ & Uniform & 0.4--0.8 & 0.5--0.95 & 0.1--0.4 \\
\midrule
\multicolumn{5}{l}{\textit{Atmospheric, aerodynamic and ablation parameters}} \\
Wind scaling & Uniform & 0.85--1.15 & 0.85--1.15 & 0.85--1.15 \\
$C_d$ scaling & Uniform & 0.85--1.15 & 0.85--1.15 & 0.85--1.15 \\
Ablation $\sigma_{\rm abl}$ & Uniform & $7$--$9 \times 10^{-8}$ kg\,J$^{-1}$ & $7$--$9 \times 10^{-8}$ kg\,J$^{-1}$ & $7$--$9 \times 10^{-8}$ kg\,J$^{-1}$ \\
\bottomrule
\end{tabularx}
\caption{Monte Carlo parameter distributions for ordinary chondrites (OC), carbonaceous chondrites (CC), and iron meteoroids. 
Density and albedo ranges are based on laboratory measurements summarized by \cite{2019P&SS..165..148O}, 
while the bimodal OC strength distribution is adopted from \cite{2020AJ....160...42B}. 
All remaining parameters are treated as modeling and uncertainty terms and sampled within conservative ranges.}
\label{tab:physical_parameters}
\end{table}

\paragraph{Initial position and velocity}
The entry state at 100~km altitude above the WGS84 ellipsoid is obtained from ESA’s Aegis software \citep{fenucci-etal_2024b}, which determines the heliocentric orbit from astrometric observations and propagates the solution to atmospheric entry. The associated impact corridor is computed using a semi-analytical method \citep{dimare-etal_2020}. The tool provides $1\sigma$, $3\sigma$, and $5\sigma$ confidence contours for latitude, longitude, velocity, flight-path angle, and heading. From these outputs, mean values and full covariance matrices for position and velocity are reconstructed, and each simulation samples an initial state from the corresponding multivariate Gaussian distribution.

\paragraph{Physical properties}
Physical properties of the meteoroid are modelled with a-priori distributions. 
The parameter ranges adopted by default are for ordinary chondrites, which account for approximately 85\% of recovered falls \citep{1996MNRAS.283..551B}, as well as for carbonaceous and iron meteoroids. The same a priori framework can be extended to other material classes if required.

  The ranges for density and albedo are based on laboratory measurements, which are summarized in \citet{2019P&SS..165..148O}. 
  The initial size of the impacting body is inferred from pre-impact photometric observations, which allow the absolute magnitude $H$ to be determined as a measure of the object’s intrinsic brightness. Converting $H$ to a physical diameter requires an assumption about the albedo $p$, which depends on the surface optical properties and composition. In the Monte Carlo framework, the albedo is sampled from a triangular distribution, with bounds and mode selected according to the assumed material type. The equivalent spherical diameter $D$ is computed using the standard photometric relation from \cite{bowell-etal_1989, 2007Icar..190..250P}
\begin{equation}
D = \frac{1329}{\sqrt{p}}\,10^{-H/5} \;\; \mathrm{km}.
\end{equation}
Although uncertainty in $H$ could be included, it is neglected here because the albedo is the dominant source of error in the inferred diameter.
Assuming a spherical shape and a uniform bulk density $\rho$, the corresponding initial mass $m_0$ is obtained as $m_0 = \pi\,\rho\,D^3/6$.

  The strength of the pre-impact material may differ substantially from that measured in recovered meteorites. For ordinary chondrites, \citet{2020AJ....160...42B} analysed seven camera-observed falls and found that the pre-entry material occupies two distinct strength ranges, roughly near $S \sim 0.08$~MPa and $S \sim 3$~MPa. These intervals are listed in Table~\ref{tab:physical_parameters}, and the corresponding a priori distribution is taken to be bimodal uniform, following \citet{2020AJ....160...42B}.
  For carbonaceous chondrites and iron meteoroids, a similar observational constraint is not available. The adopted strength ranges are therefore selected to reflect the expected mechanical behaviour of these materials: lower cohesive strengths for carbonaceous chondrites and significantly higher strengths for iron bodies, consistent with laboratory measurements and typical fragmentation altitudes reported in the literature. In these cases, a uniform distribution is assumed within conservative bounds following \citet{2020AJ....160...42B}.

\paragraph{Fragmentation parameters}
Ordinary chondrites display structural properties intermediate between
weak, porous carbonaceous meteoroids and strong, coherent iron bodies.
Their chondrule–matrix texture, moderate porosity, and collisional
fracturing typically produce breakup behaviour that is neither extremely
dust-rich nor dominated by a single surviving mass. The adopted dust fraction (0--0.9), largest fragment fraction ($m_l=0.05$--0.4), and $\beta$ range (0.4--0.8) reflect this intermediate behaviour, allowing for multiple fragments without extreme dispersion.
Carbonaceous chondrites are assumed to fragment more efficiently due to their higher porosity and lower cohesion. Accordingly, higher dust fractions (0.3--0.95), smaller largest-fragment fractions ($m_l=0.01$--0.25), and larger $\beta$ values (0.5--0.95) are adopted to represent more dispersive outcomes.
Iron meteoroids, being mechanically stronger and more cohesive, are expected to retain a dominant main mass. For this reason, low dust fractions (0.0--0.25), large $m_l$ values (0.4--0.9), and smaller $\beta$ (0.1--0.4) are assumed.

\paragraph{Atmospheric, aerodynamic and ablation parameters}
To account for uncertainties in the GFS wind modelling, we assume a 15\% error on the data retrieved for the impact epoch and location. Therefore, the profile is rescaled by a random factor ranging from 0.85 to 1.15, which we assume to be uniformly distributed. This assumed error value is conservative for impacts predicted up to 2--3 days in advance, as GFS data is expected to achieve better performance on shorter prediction times.
A similar scaling range (0.85--1.15) is adopted for the drag coefficient $C_d$ to account for uncertainties in the aerodynamic formulation, including deviations from the idealized assumptions of the Ceplecha-type modelling and the effects of non-spherical shape and attitude variations during flight.
The ablation coefficient $\sigma_{\rm abl}$ is sampled uniformly in the range $7$--$9 \times 10^{-8}$ 
kg\,J$^{-1}$, a value empirically determined in this work through numerical tests 
reproducing the strewn fields and recovered masses of observed falls (the NEAs 2008~TC3, 
2018~LA, 2022~WJ1, 2023~CX1, 2024~BX1, and the Chelyabinsk, Winchcombe, Cavezzo, Antonin, 
Golden and Madura Cave meteorites) as reported in the Meteoritical Bulletin \citep{MetBull}.

\section{Operational integration within the Meerkat/Aegis pipeline}
\label{s:meerkataegis}
  The strewn field prediction framework presented is not conceived as a stand-alone numerical model, but as an operational component integrated within ESA’s imminent impactor monitoring infrastructure operated by the NEO Coordination Centre (NEOCC). Its development was motivated by the increasing frequency of meteoroids discovered only hours before atmospheric entry, for which rapid predictions are required.

  The operational workflow connects Meerkat, Aegis, and the present strewn field solver within a fully automated chain. The whole pipeline scheme is summarized in the flowchart of Fig.~\ref{fig:flow_diagram}. Meerkat \citep{drury-etal_2026} continuously monitors the NEO Confirmation Page (NEOCP) and applies systematic ranging \citep{farnocchia-etal_2015c} to unconfirmed NEAs, sampling the admissible orbital region to assess impact probability in near real time. When an impact score larger than 10\% is identified for an object having more than 4 observations, Aegis takes over the orbit determination process, refining the solution through differential correction and covariance analysis, and computing the impact corridor \citep{gianotto-etal_2025}. If the impact probability found is 100\% and the $1\sigma$ uncertainty is smaller than 10 km, the impact corridor is passed directly to this module. Specifically, Aegis provides the strewn field software with the 
impact corridor defined by the $1\sigma$, $3\sigma$, and $5\sigma$ 
confidence contours for latitude, longitude, velocity, flight-path angle, 
and heading, together with an orbit file containing the impactor name and 
absolute magnitude $H$. The strewn field computation is implemented in \texttt{python} and parallelized through multiprocessing to handle large Monte Carlo simulations. The results are finally sent by email to NEOCC operators. As new astrometric observations are acquired in real time and submitted to the NEOCP, the pipeline is automatically re-executed, producing progressively refined impact predictions. The improvement originates at the orbit determination level: additional measurements reduce the extent of the admissible orbital region and shrink the semi-major axis of the positional uncertainty ellipse at 100 km altitude from ground. This directly translates into a smaller uncertainty in the initial conditions passed to the strewn field solver.

  The outputs of the Monte Carlo model consist of probabilistic products generated before atmospheric entry, or until the impacting object stays on the NEOCP. These include a static two-dimensional impact probability map and an interactive HTML\footnote{\url{https://doi.org/10.5281/zenodo.18803957}} version allowing dynamic exploration of the results. The latter enables switching between impact probability, cumulative deposited mass, and discrete mass bins, with additional controls for contours, transparency, and spatial overlays. Complementary summary plots provide statistical distributions of key quantities (e.g., number of surviving fragments, total mass at ground, breakup altitude, number of breakup events, ablated fragments, time of flight, and impact speed), together with the atmospheric wind profile used during propagation.

  All products can be immediately used by downstream services without additional post-processing. This enables fast operational use, including coordination with civil protection authorities and pre-positioning of meteorite recovery teams. The framework operates independently from all-sky camera networks for fireball observations, but it remains fully compatible with their later inclusion for post-entry refinement \citep{2024PSJ.....5..253K, 2025NatAs...9.1624E}.

\begin{figure}
    \centering
    \includegraphics[width=0.3\linewidth, angle=90]{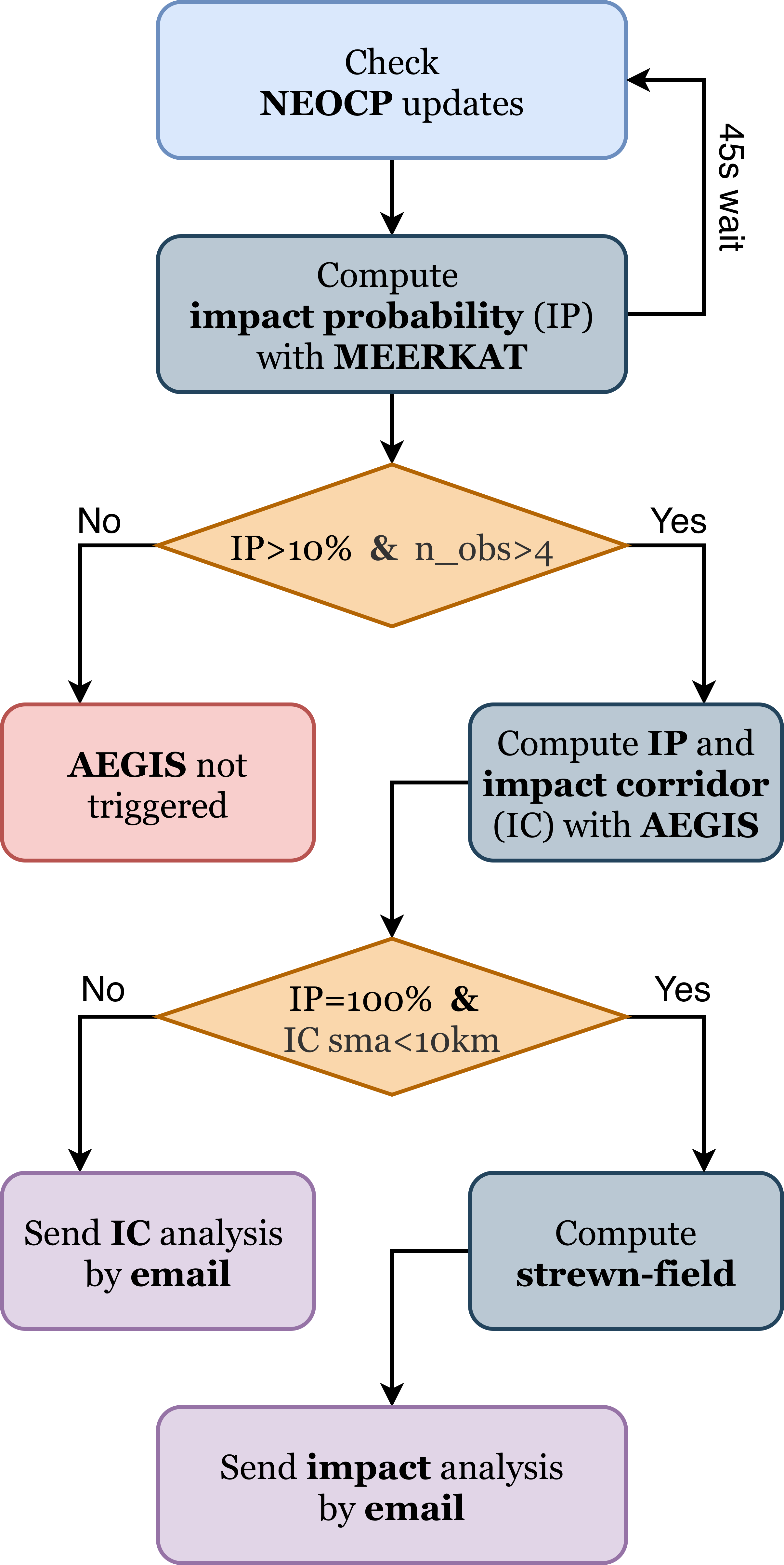}
    \caption{Operational pipeline for imminent impactors. After NEOCP monitoring, 
impact probability (IP) is computed with Meerkat. If IP $>$ 10\% and the number 
of observations $N_{\mathrm{obs}} > 4$, the impact probability and corridor (IC) 
are derived with Aegis. Once IP = 100\% and the semi-major axis uncertainty 
($\sigma_a$, sma) is below 10 km, atmospheric entry and fragmentation simulations 
are performed and operational products are automatically generated and made 
available without post-processing.}
    \label{fig:flow_diagram}
\end{figure}

\section{Tests on previous impactors}
\label{s:results}
\subsection{2023~CX1}

  The fall of asteroid 2023~CX1 occurred on 13 February 2023, when the metre-sized object entered Earth’s atmosphere over northern France at approximately 02:59~UTC and disintegrated over the Normandy region \citep{2025NatAs...9.1624E}. It was discovered less than seven hours before impact and predicted to strike above the English Channel between 02:00 and 04:00~UTC \citep{2025NatAs...9.1624E}. The event produced a bright fireball visible across multiple countries, and coordinated search efforts led to the recovery of multiple fragments in Normandy. The first one, weighing about 93~g and classified as L-type ordinary chondrite, was found near Saint-Pierre-le-Viger and has been named after it. This case represents an ideal test case to validate the 
full Meerkat/Aegis operational pipeline in absence of new pre-impact detections since December 2024. 
This event benefits from well-constrained orbital data, a set of recovered meteorites providing ground truth, and numerous fireball 
observations available for independent verification.

  To reproduce the operational conditions, we reconstructed the 
sequence of observations as they were progressively posted on the NEOCP, 
computing orbital solutions at roughly 30 minutes intervals during the discovery 
arc. Observational data as announced on the NEOCP are available at the Minor Planet Center\footnote{\url{https://minorplanetcenter.net/cgi-bin/cgipy/dblog?desig=Sar2667}}.

  As illustrated in Fig.~\ref{fig:sma}, increasing the number of optical observations systematically reduces the entry-state uncertainty. The 1$\sigma$ in the impact point at 100 km ranges from $\sim9$~km at 4.6~h before impact, to below 100~m at approximately 0.8~h. This rapid reduction directly translates into a progressive narrowing of the predicted impact region.

  The initial state adopted for the final simulations is reported 
in Table~\ref{tab:cx1_init}, corresponding to the best pre-impact orbit 
solution, with a semi-major axis uncertainty of only 49~m. The atmospheric profile used corresponds to 
03:00~UTC (Fig.~\ref{fig:wind}), i.e. very close to the impact epoch, ensuring 
a highly accurate representation of the atmospheric state during the 
event. The same wind profile is used for all runs. Small variations of it are included in the Monte Carlo sampling, while the overall vertical structure of the profile is kept fixed.

\begin{table}[h]
\centering
\begin{tabular}{lc}
\hline
\textbf{Parameter} & \textbf{2023~CX1}\\
\hline
Time (UTC) & 2023-02-13 02:59:13.387 \\
Latitude (deg) & $49.919022 \pm 0.00012$ \\
Longitude (deg) & $-0.1467625 \pm 0.0006545$ \\
Speed (km s$^{-1}$) & $14.01645 \pm 0.000185$ \\
Flight path angle (deg) & $-49.153055 \pm 0.00051$ \\
Heading angle (deg) & $101.386628 \pm 0.000393$ \\
1-$\sigma$ semi-major axis (m) & 49 \\
1-$\sigma$ semi-minor axis (m) & 5.2 \\
1-$\sigma$ azimuth (deg) & 105.48 \\
\hline
\end{tabular}
\caption{Impact parameters at 100 km altitude (WGS84 ellipsoid) for asteroid 2023~CX1 as computed by Aegis. The velocity is given with respect to the ground.}
\label{tab:cx1_init}
\end{table}

\begin{figure}
    \centering
    \includegraphics[width=0.6\linewidth]{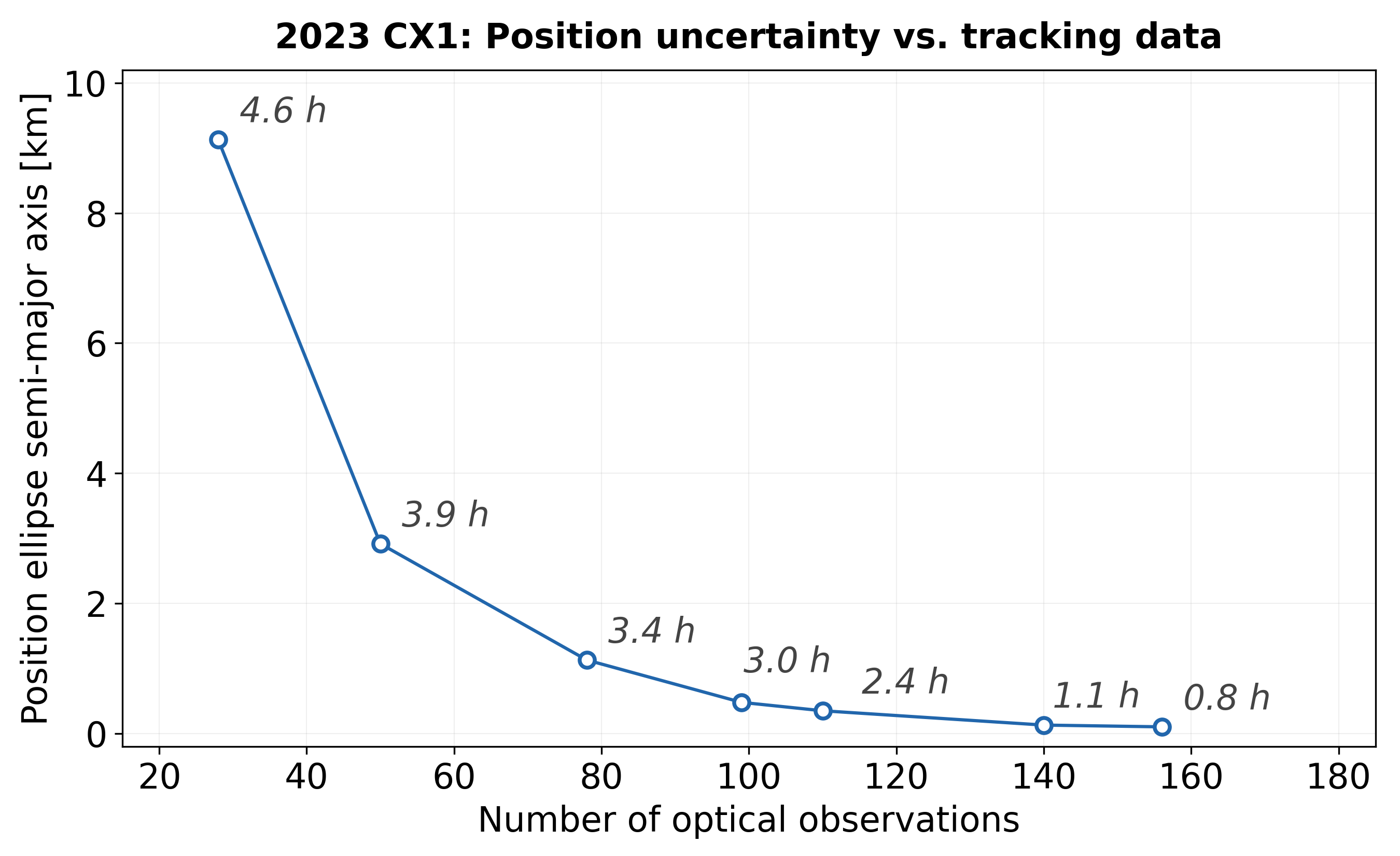}
\caption{Semi-major axis of the position uncertainty ellipse at 100 km altitude versus the number of optical observations for 2023~CX1, with labels indicating time before impact. Earlier times (left) correspond to fewer observations and larger uncertainty, while additional tracking closer to impact progressively reduces the predicted position uncertainty.}
    \label{fig:sma}
\end{figure}
\begin{figure}
    \centering
\includegraphics[width=0.7\linewidth]{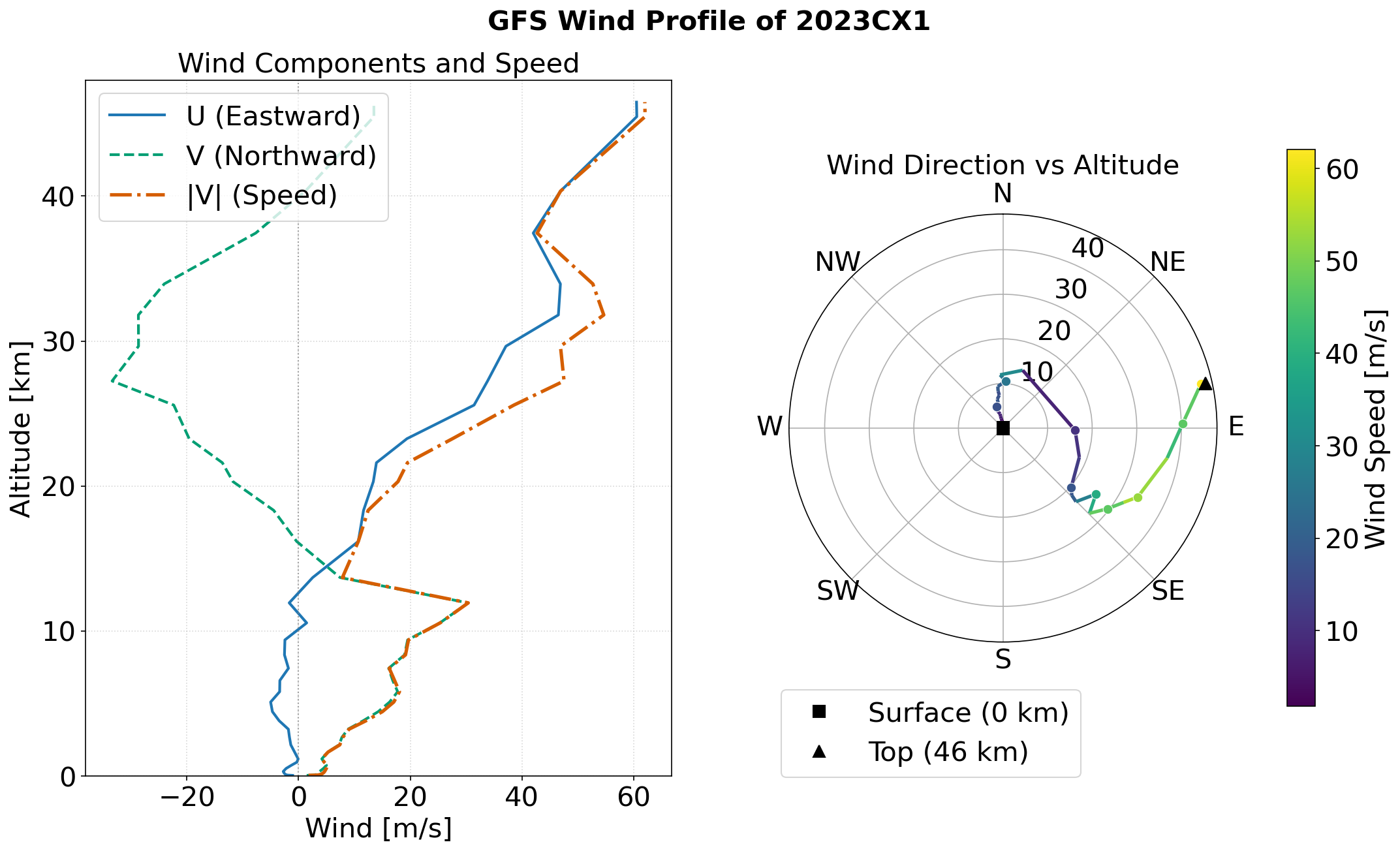}
    \caption{GFS-derived atmospheric wind profile for the 2023 CX1 event. Left: zonal (U, eastward), meridional (V, northward), and total wind speed as a function of altitude. Right: wind direction versus altitude in polar representation, with color indicating wind speed; markers denote surface and top-of-profile levels.}
    \label{fig:wind}
\end{figure}

  The evolution of the predicted impact region is shown in Fig.~\ref{fig:kdecx1_comparison}. At early 
stages of the tracking campaign the probability region is broad and 
roughly oval, with little indication of the final strewn field 
orientation. As additional observations are incorporated, the predicted 
ground distribution progressively shrinks and rotates, converging toward 
the final geometry and orientation of the actual strewn field. Notably, already 3.9 hours before impact the predicted probability region reproduces the correct orientation.

  The final Monte Carlo results are shown in Fig.~\ref{fig:kdecx1_comparison} 
in the form of an impact probability map. The predicted distribution 
reproduces well the position, orientation, and spatial extent of the 
observed strewn field. All but one recovered meteorite (out of 93) fall within the 95\% density contour,
and all are contained within the 99\% contour.
The meteorite recovery locations were obtained from the LPI Strewn Field database \citep{lpi_cx1}.

  It is worth noting that the probability map shows an extended high-probability region where no meteorites have been recovered. This feature arises from the way the Kernel Density Estimation (KDE) map is constructed, as it represents the spatial density of simulated impacts rather than being weighted by fragment mass. During fragmentation, a large number of small bodies are generated, and they typically carry larger trajectory uncertainties than the heavier ones. As a result, their predicted impact points are more dispersed, producing a broader and more elongated probability trail that extends beyond the locations of the recovered meteorites. Moreover, many of the smallest fragments are unlikely to be found in the field, either because they are too small to be easily detected or because they were not systematically searched for, so their potential presence may not be reflected in the recovery data. Conversely, the largest recovered meteorites, visible toward the downrange end of the strewn field, correspond to relatively rare high-mass impacts; since this map counts impacts rather than weighting them by mass, these events contribute less to the overall probability density and therefore lie in darker, lower-probability regions. A mass-weighted probability map instead emphasize these larger fragments and produce higher probability values in their vicinity, like in Fig.~\ref{fig:kdemasscx1}.

  Within the Monte Carlo ensemble, individual simulations exhibit 
a wide range of physically plausible trajectories. Among these, one run 
provides an especially close match to the recovered meteorites 
(Fig.~\ref{fig:nominalcx1}). For this representative case, the median 
distance between simulated impact points and recovered meteorite 
locations is 151~m, highlighting the level of accuracy that can be achieved when the model is properly tuned. The example discussed is intended only to illustrate the level of agreement that can be reached by one physically plausible realization when additional information is available, for example after the event or from fireball observations. The parameters used 
for this run are: $E_0 = 3.6878\times10^{-5}$ kt\,TNT, density 
$\rho = 3000$~kg\,m$^{-3}$, strength $S = 4$~MPa, ablation parameter 
$\sigma_{\rm abl} = 8\times10^{-8}$~kg\,J$^{-1}$, dust fraction $=0.6$, $\beta=0.55$, 
and largest fragment mass fraction $m_l=0.2$. These values are in good 
agreement with the fireball analysis of \citet{2025NatAs...9.1624E}: in 
particular, the adopted density and strength are consistent with those 
reported therein, and the model produces a single main fragmentation event 
at $\sim$28~km altitude, matching the point identified 
observationally.

  Fig.~\ref{fig:fragtimecx1} and ~\ref{fig:fragcx1} illustrates the time evolution and trajectories of the fragments after the main single breakup for the best Monte Carlo realisation providing the closest match to the recovered meteorites. Smaller fragments are more strongly affected by aerodynamic drag and winds, which keep them aloft for a longer time, so they take longer to reach the ground and generally land closer to the breakup point. Larger fragments, being less influenced by drag, descend more directly and tend to achieve a larger ground range. Note that the dynamic pressure continues to increase after breakup because the atmospheric density initially increases faster than $|\mathbf{V}_r|^2 $ decreases. Breakup occurs when $p_{\rm ram}=\frac{1}{2}\rho_{\rm air}|\mathbf{V}_r|^2$ reaches the parent strength, while the stronger child fragments survive the subsequent pressure increase. For 2023~CX1, $p_{\rm ram}$ reaches its maximum before the rapid deceleration below approximately 23~km causes it to decrease.

\begin{figure}
    \centering
    \includegraphics[width=0.8\textwidth]{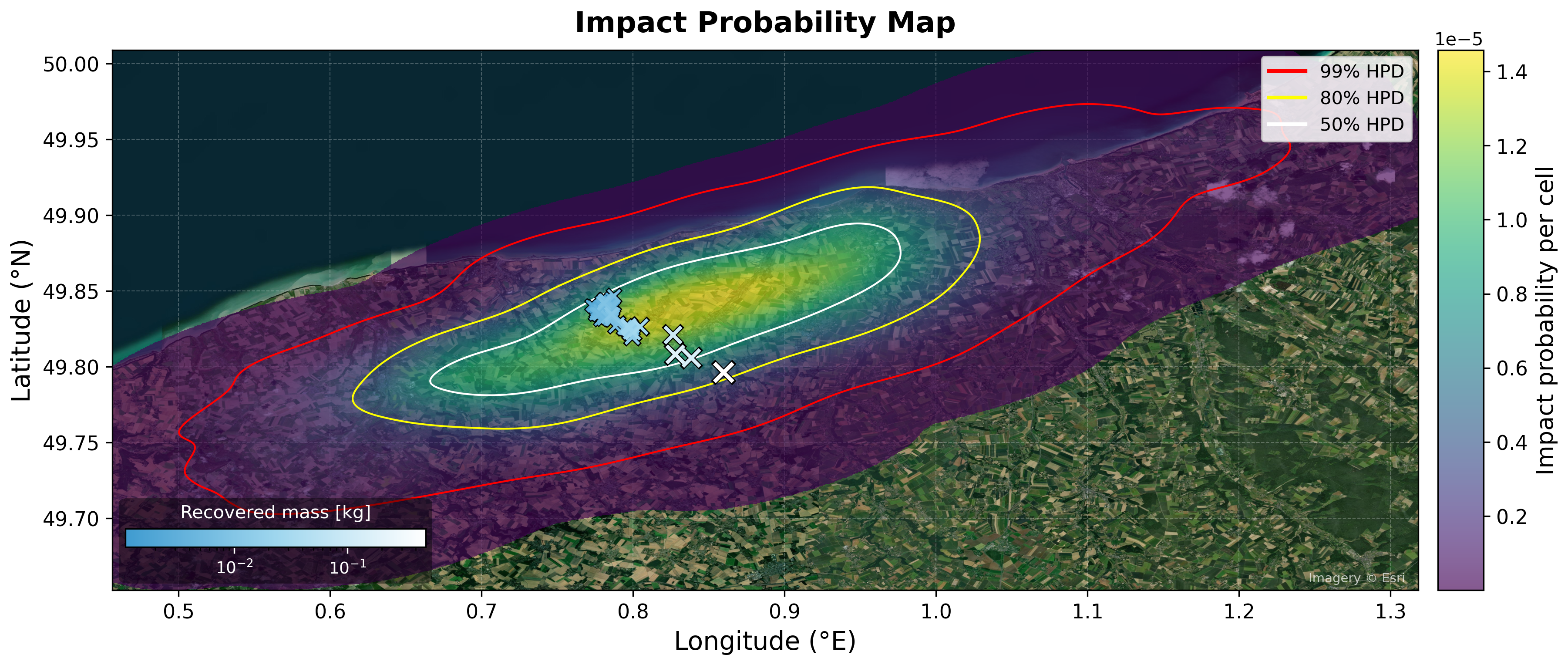}

    \vspace{0.5cm} % spazio opzionale

    \includegraphics[width=0.8\textwidth]{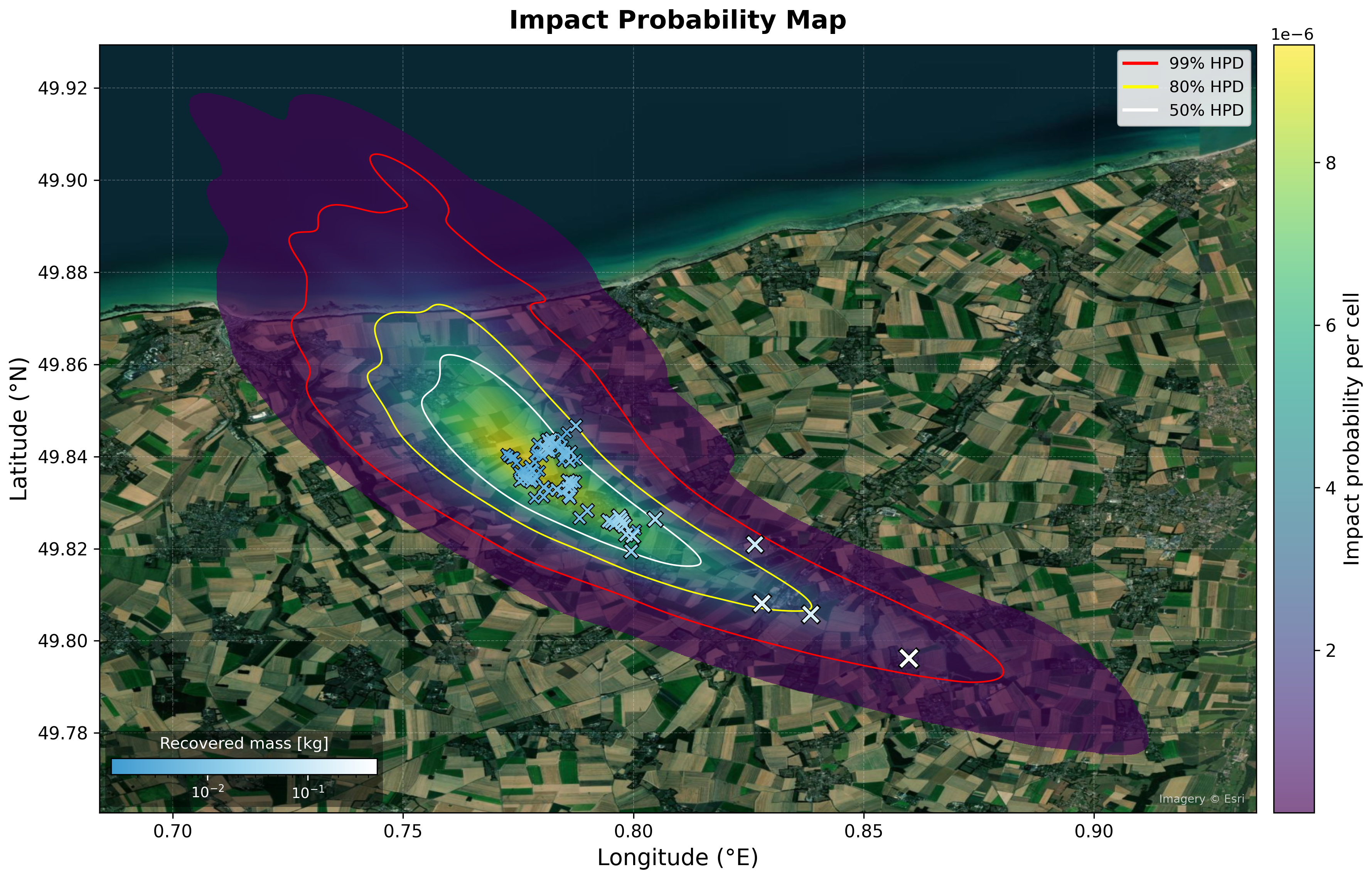}

    \caption{KDE-based impact probability maps for 2023 CX1. Top: earliest solution (4.6 h before impact), corresponding to the largest positional uncertainty. Bottom: latest solution (0.8 h before impact), corresponding to the smallest positional uncertainty. Colors represent the spatial impact probability density, contours denote the highest posterior density regions, and crosses mark recovered meteorites (size and color proportional to mass).}
    \label{fig:kdecx1_comparison}
\end{figure}

\begin{figure}
    \centering
    \includegraphics[width=1\linewidth]{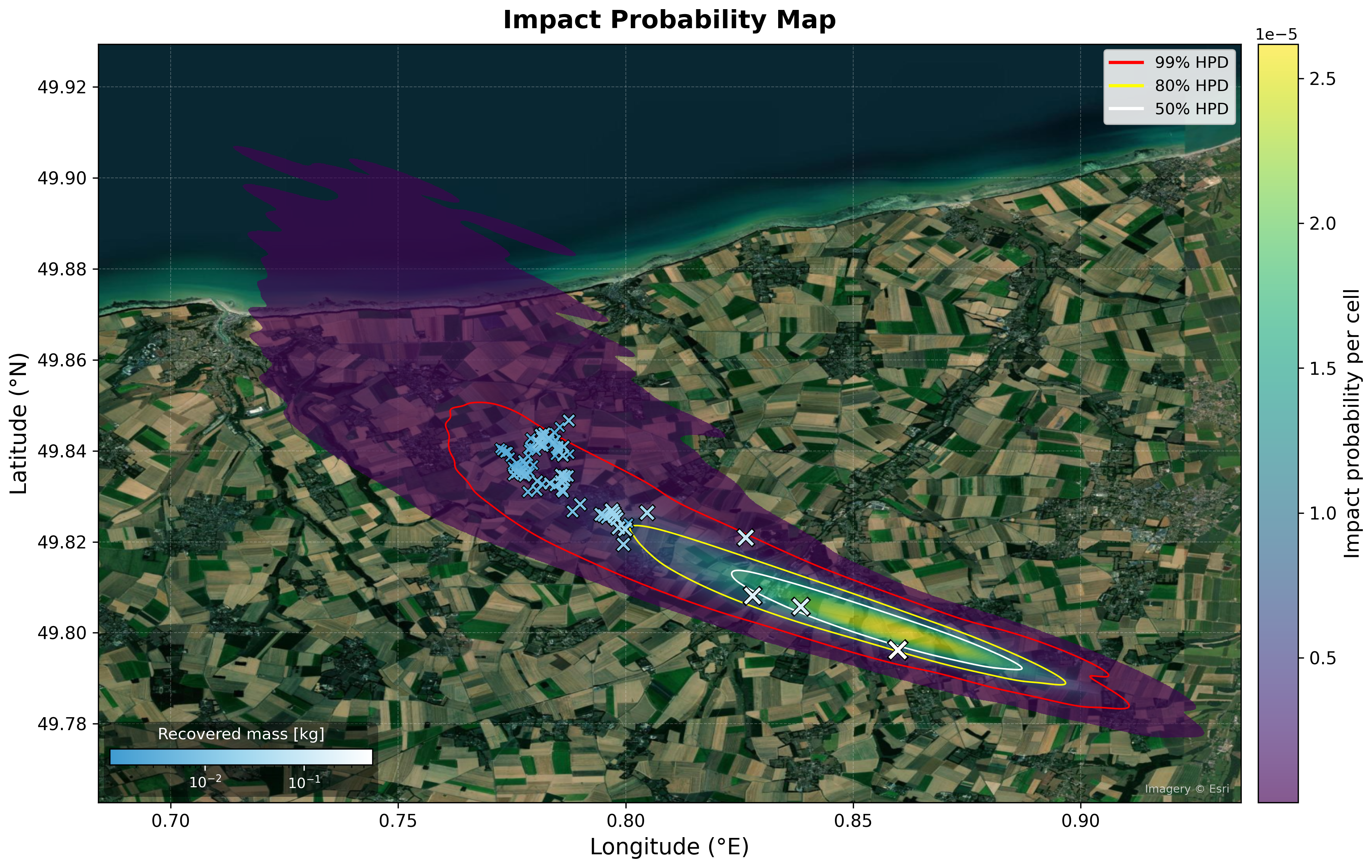}
    \caption{KDE-based mass probability map for the latest solution (0.8 h before impact) of 2023 CX1, corresponding to the smallest positional uncertainty. Colors represent the spatial mass probability density, contours denote the highest posterior density regions, and crosses mark recovered meteorites (size and color proportional to mass).}
    \label{fig:kdemasscx1}
\end{figure}
\begin{figure}
    \centering
\includegraphics[width=0.7\linewidth]{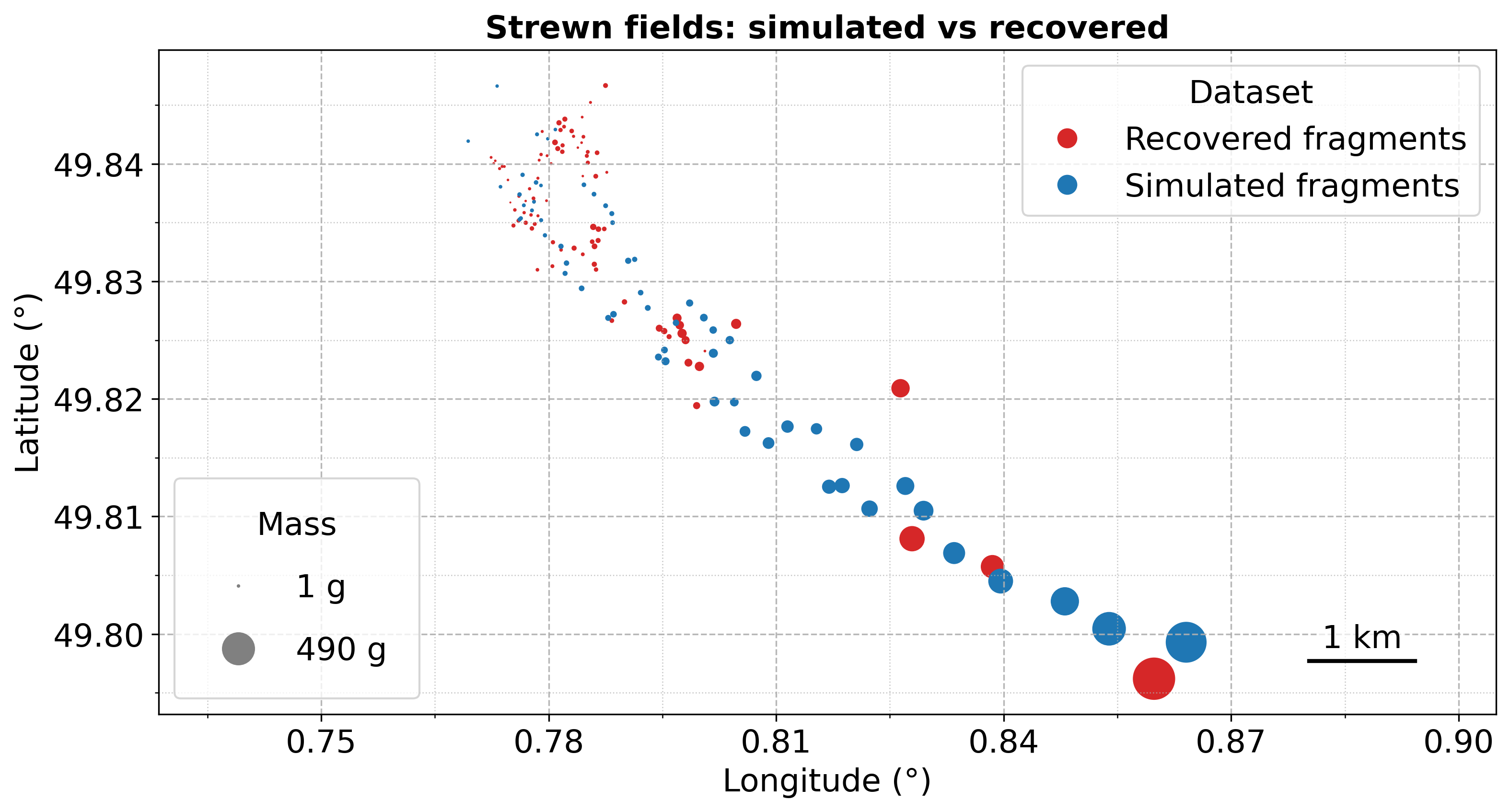}
\caption{Monte Carlo realisation providing the closest match to the recovered meteorites for 2023~CX1. Simulated fragment impacts are shown together with recovered meteorite locations; marker size is proportional to fragment mass.}
    \label{fig:nominalcx1}
\end{figure}
\begin{figure}
    \centering
    \includegraphics[width=0.7\linewidth]{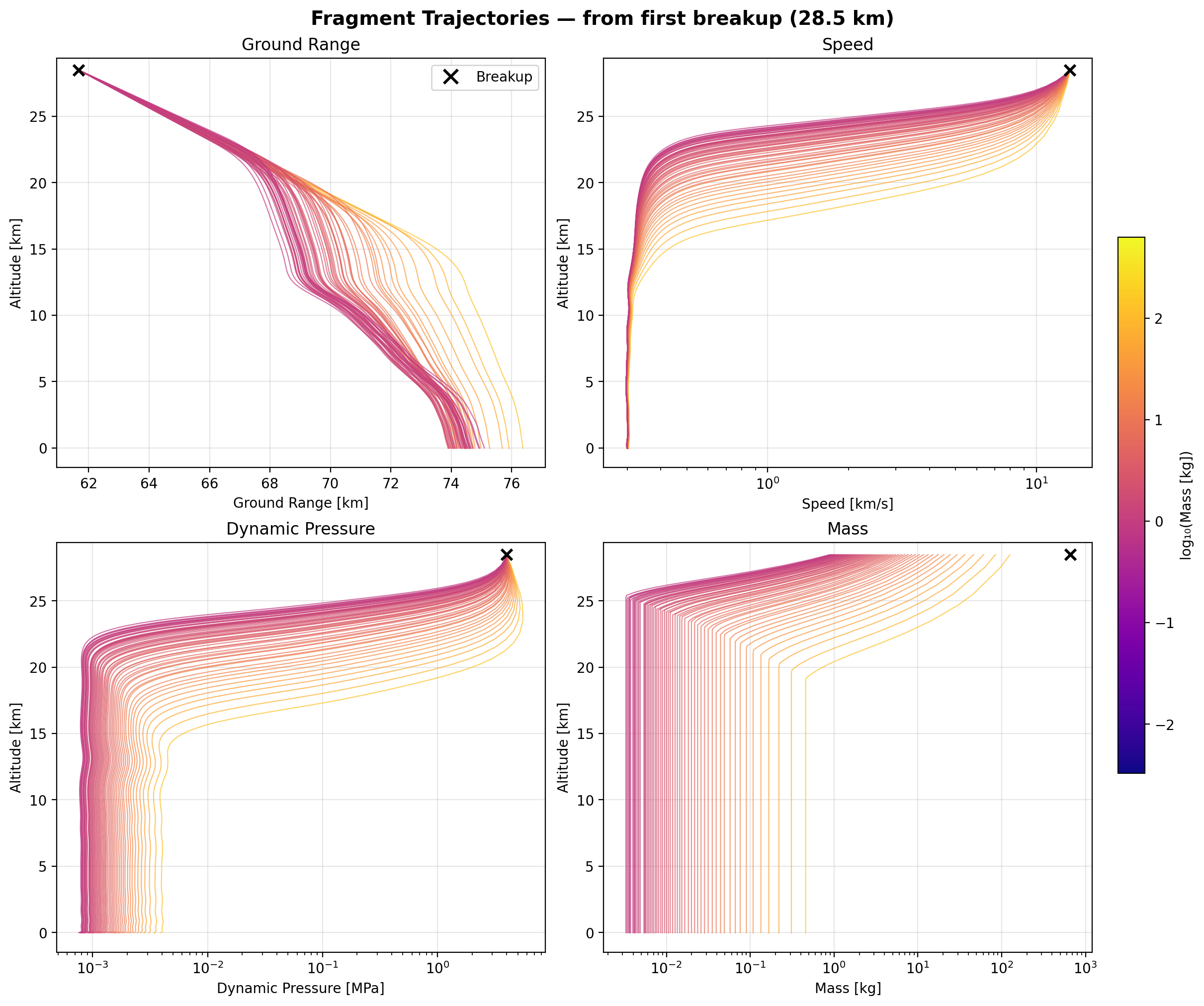}
\caption{Fragment time-series evolution for meteoroid 2023~CX1 for the best Monte Carlo realisation providing the closest match to the recovered meteorites. The plots show altitude, ground range, speed, and mass, with the breakup point marked by X and colors indicating fragment mass. The event is characterized by a single main fragmentation episode rather than multiple successive breakups. Smaller fragments remain airborne longer and generally land closer to the breakup point, while larger fragments tend to travel farther and impact at greater ground range}
    \label{fig:fragtimecx1}
\end{figure}
\begin{figure}
    \centering
    \includegraphics[width=0.7\linewidth]{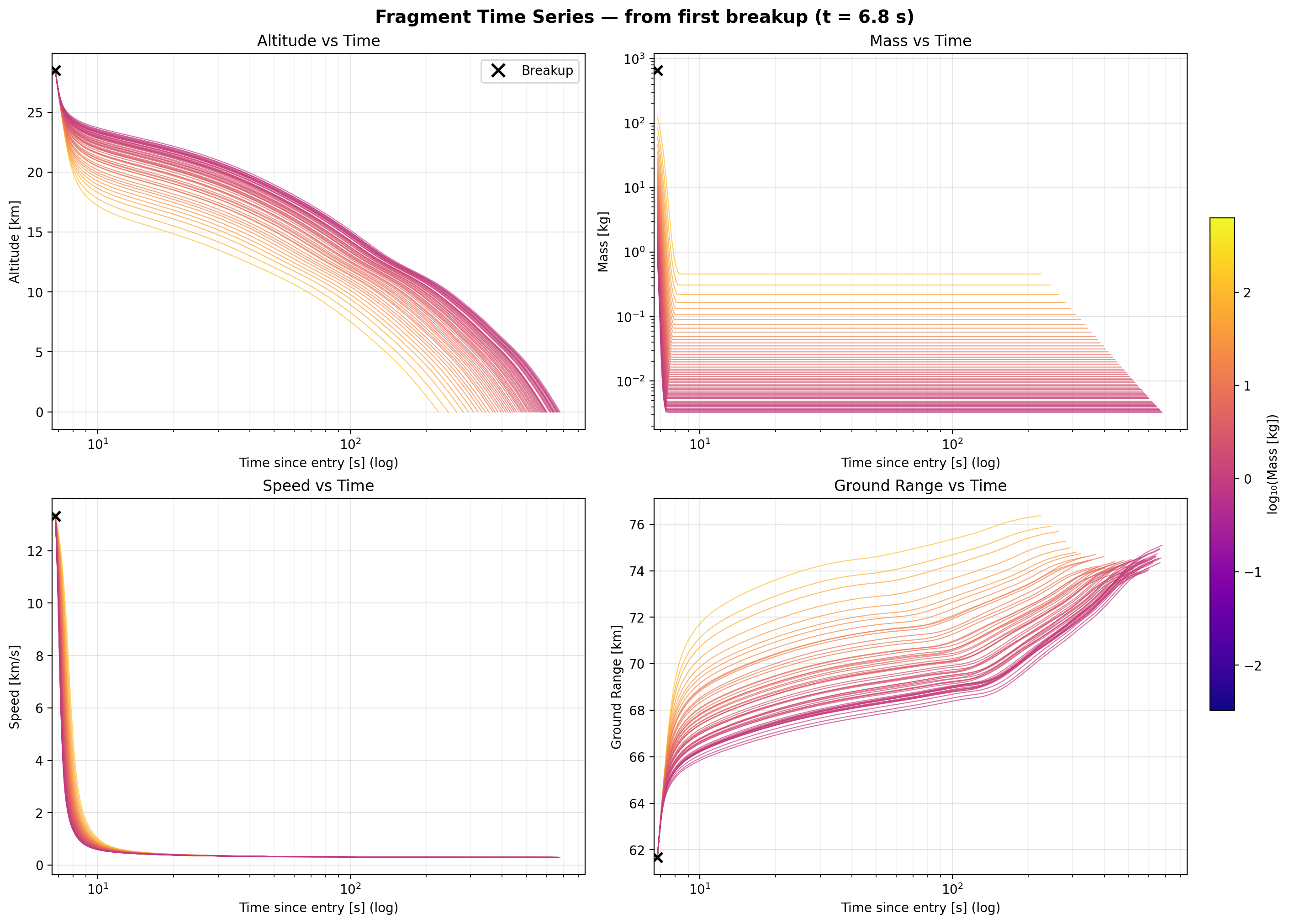}
\caption{Fragment trajectories evolution for meteoroid 2023~CX1 for the best Monte Carlo realisation providing the closest match to the recovered meteorites. The plots show altitude, ground range, speed, dynamic pressure, and mass, with the breakup point marked by X and colors indicating fragment mass.}
    \label{fig:fragcx1}
\end{figure}

\subsection{2008~TC3}

  The 2008~TC3 event occurred on 7 October 2008 when a small asteroid entered Earth’s atmosphere above the Nubian Desert in northern Sudan. It was the first asteroid discovered prior to impact, 20 hours before atmospheric entry \citep{2010M&PS...45.1557S, farnocchia-etal_2017}, and its pre-impact orbit was sufficiently well constrained to define a narrow impact corridor, as in Table \ref{tab:tc3_init}. No ground-based fireball triangulations were available for this event \citep{Borovicka2009}, thus the atmospheric trajectory reconstruction relied primarily on satellite detections and eyewitness reports. For this reason, 2008~TC3 falls squarely within the class of impacts for which an \emph{ab initio} strewn field prediction, based solely on orbital data and atmospheric modeling, becomes essential.

\begin{table}[h]
\centering
\begin{tabular}{lc}
\hline
\textbf{Parameter} & \textbf{2008~TC3}\\
\hline
Time (UTC) & 2008-10-07 02:45:30.09 \\
Latitude (deg) & $21.0884 \pm 0.0009$ \\
Longitude (deg) & $30.5347 \pm 0.0038$ \\
Speed (km s$^{-1}$) & $12.38041  \pm 0.00005$ \\
Flight path angle (deg) & $-20.8360 \pm 0.0030$ \\
Heading angle (deg) & $101.0953 \pm 0.0015$ \\
1-$\sigma$ semi-major axis (m) & 590.3 \\
1-$\sigma$ semi-minor axis (m) & 38.1\\
1-$\sigma$ azimuth (deg) & 104.75\\
\hline
\end{tabular}
\caption{Impact parameters at 100 km altitude (WGS84 ellipsoid) for asteroid 2008~TC3. The velocity is given with respect to the ground.}
\label{tab:tc3_init}
\end{table}

  Numerous fragments were subsequently recovered over a widespread strewn field, with the collected material named the Almahata Sitta meteorites \citep{Borovicka2009}, which were a polymict ureilite (a rare achondritic meteorite type), providing an exceptionally rich dataset for validation. The recovered fragments exhibit a distinct variation in cross-track dispersion along the strewn field axis. In particular, smaller fragments are observed to cluster more tightly than larger ones in certain sections of the field. This behaviour is not attributable to wind variability but rather to the fragmentation dynamics during atmospheric entry. Fragments released during an earlier high-altitude flare were likely confined within the wake region of the parent body, resulting in a more compact lateral distribution \citep{Jenniskens2022}. In contrast, major fragments produced during the main fragmentation episode at lower altitude were ejected with higher relative velocities, leading to a broader cross-track spread.

  The Monte Carlo does not resolve fine-scale
fragmentation physics (e.g., wake confinement), since breakup is modeled phenomenologically.
Accordingly, it does not reproduce the observed local clustering of
fragments. It nevertheless captures the overall extent and orientation
of the strewn field, with all recovered meteorites falling within the
predicted probability region. Compared to the deterministic solution of
\citet{2025Icar..42516345C}, this provides a more realistic
representation of the impact distribution.

  Since the NEOCP system was not yet in place at the time of the
event, this test was performed using the final reconstructed orbit only.
The adopted solution is based on the complete observational data set
available at the MPC and is characterized by a semi-major axis
uncertainty of about 590~m. Although this is less precise than the
$\sim$49~m uncertainty reached for 2023~CX1 shortly before impact, it is
still well below the kilometre level and therefore provides a highly
constrained entry state.

  The Monte Carlo results show good agreement between the
predicted impact probability map and the observed strewn field geometry,
both in terms of orientation and spatial extent (see
Fig.~\ref{fig:kdetc3}). The simulated distribution reproduces the
elongated downrange dispersion of the fragments and the overall lateral
spread of the field. 94.7\% of recovered meteorites fall
within the 95\% probability region, and all are contained within the
99\% probability envelope, confirming that the model captures the
observed ground distribution despite the larger orbital uncertainty
compared to 2023~CX1. The meteorite recovery locations were obtained from \cite{2010M&PS...45.1557S}.

  As in the 2023~CX1 case, the Monte Carlo ensemble spans a wide
range of physically admissible fragmentation scenarios. Among these, one
representative run reproduces the recovered meteorite distribution with
high spatial accuracy, with a median offset of 138~m from the observed
fall locations (Fig.~\ref{fig:nominaltc3}). The parameters adopted for
this run are: diameter $D=3.6$~m, density $\rho=2000$~kg\,m$^{-3}$,
strength $S=0.5$~MPa, ablation parameter $\sigma_{\rm abl}=9\times10^{-8}$~kg\,J$^{-1}$,
dust fraction $=0.9$, $\beta=0.45$, and largest fragment mass fraction
$m_l=0.1$. The adopted density is consistent with the values reported 
by \citet{2010M&PS...45.1557S}, who measured bulk densities in the range 
$1.5$--$3.4$~g\,cm$^{-3}$.
The model produces a major fragmentation event at $\sim$42~km altitude, 
in agreement with the disruption onset reported at 46--42~km by 
\citet{2010M&PS...45.1557S}, followed by successive fragmentation episodes 
at slightly higher strengths.

\begin{figure}
    \centering
\includegraphics[width=1\linewidth]{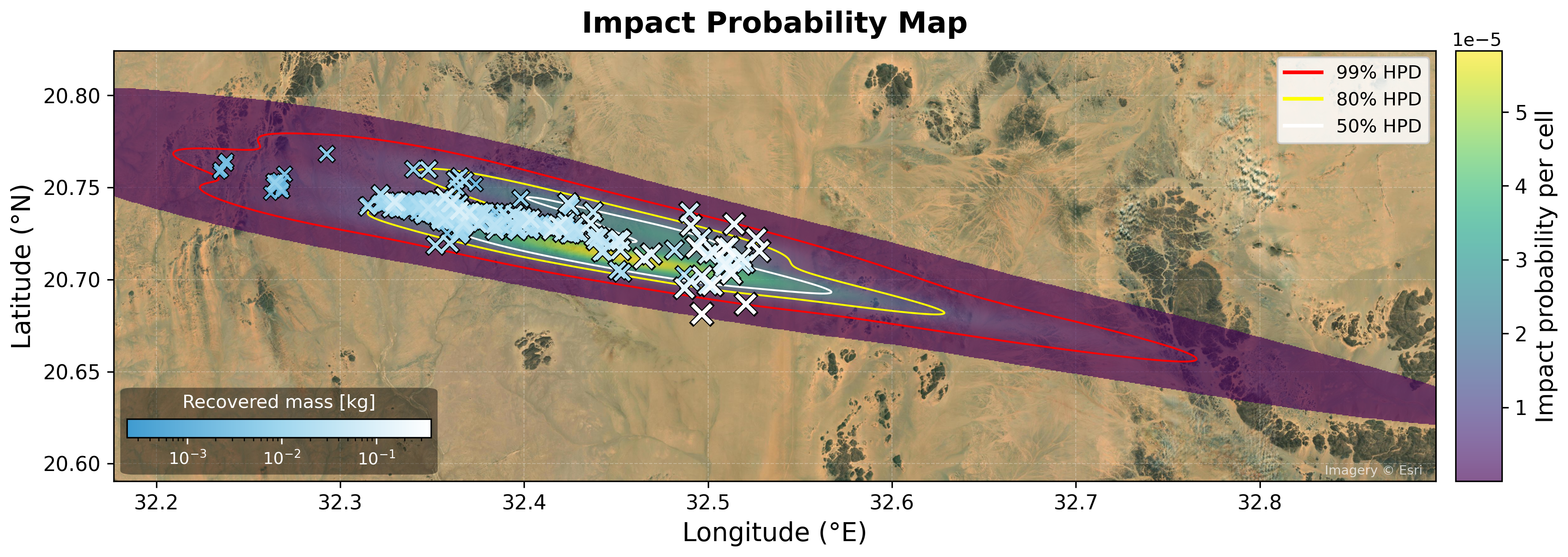}
    \caption{KDE-based impact probability map for the 2008~TC3 meteoroid. Colors show the spatial impact probability density, contours indicate the highest posterior density regions, and crosses mark recovered meteorites (size and color proportional to mass).}
    \label{fig:kdetc3}
\end{figure}
\begin{figure}
    \centering
\includegraphics[width=0.7\linewidth]{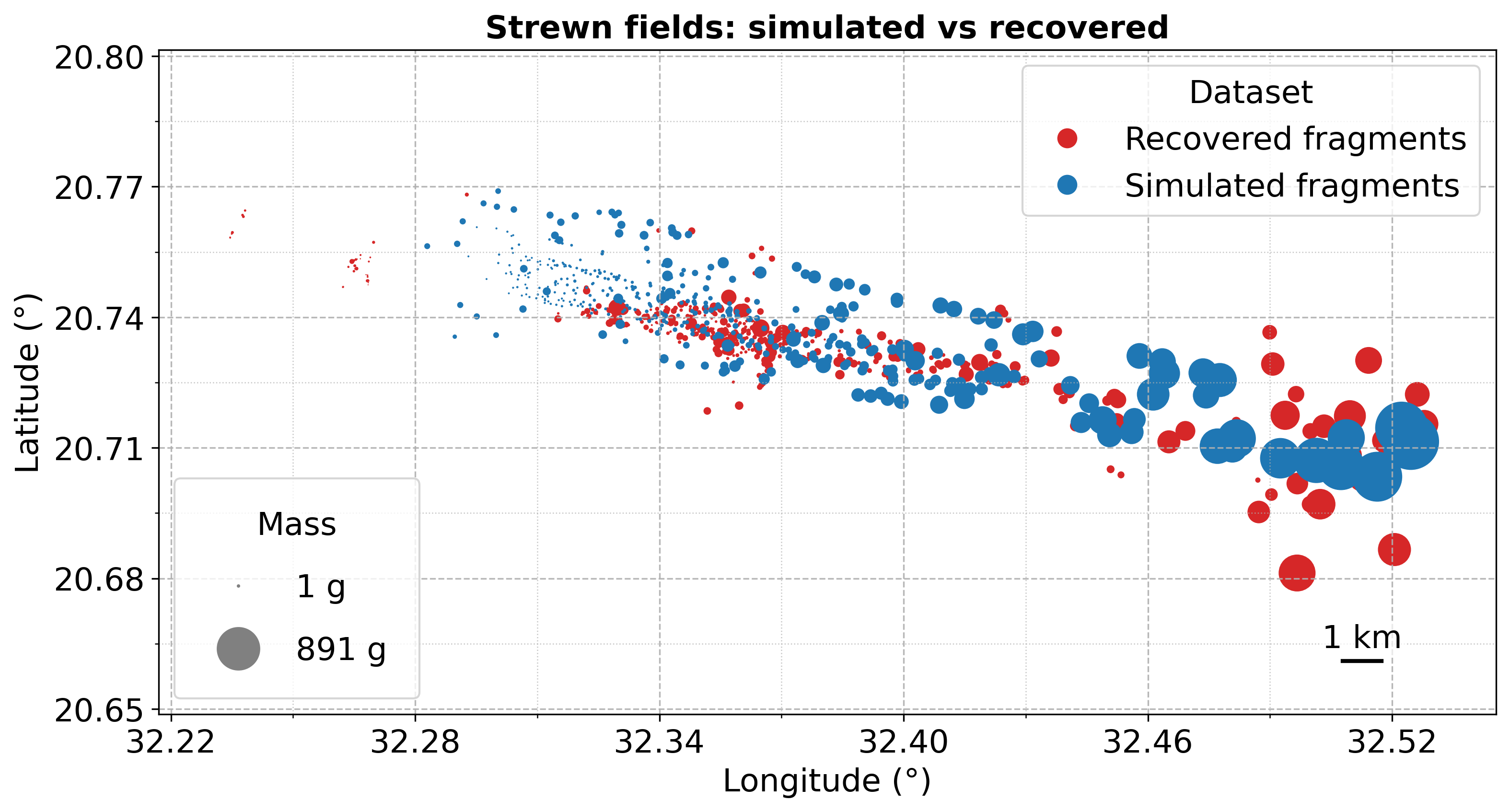}
\caption{Monte Carlo realisation providing the closest match to the recovered meteorites for 2008~TC3. Simulated fragment impacts are shown together with recovered meteorite locations; marker size is proportional to fragment mass.}
    \label{fig:nominaltc3}
\end{figure}

\subsection{2018~LA}
  Asteroid 2018~LA impacted on 2 June 2018 over Botswana with a warning time of 8.5 hours. Due to the lack of substantial astrometric data, the impact location computed from orbit determination was affected by large uncertainties. The area is also not covered by all-sky camera networks, so meteorite search campaigns were organized on the basis of triangulations made using video observations of the fireball \citep{jenniskens-etal_2021}. A total of 33 meteorites, later named Motopi Pan, were recovered, and the estimated diameter of the asteroid prior to atmospheric entry was 156 cm.

  Recently, new astrometric positions were extracted from images obtained by the SkyMapper Southern Survey (MPC observatory code Q55). Although these observations were known and already discussed in \citet{jenniskens-etal_2021}, the corresponding astrometry had not been submitted to the MPC until recently. Including this new set of measurements in the orbit determination results in a better-constrained orbit, with a 1$\sigma$ uncertainty in the impact point of 7.09 km.

  The strewn field prediction computed with our Monte Carlo method (see Fig.~\ref{fig:2018LA_sf}) with initial conditions as in Table \ref{tab:la_init} shows that all but one recovered meteorites fall within the 80\% probability area. It should be noted that a 
1$\sigma$ uncertainty of 7.09~km remains relatively large; nevertheless, 
the recovered meteorites are offset from the centre of the predicted 
distribution by only $\sim$1.67~km, which is remarkably small given the 
orbital uncertainty, and confirms that the predicted strewn field is 
sufficiently constrained to support the organisation of a recovery 
campaign. Had these additional observations been reported to the MPC at the 
time of the impact event, results from our method would have provided a 
valuable contribution to the Motopi Pan campaign.

\begin{table}[h]
\centering
\begin{tabular}{lc}
\hline
\textbf{Parameter} & \textbf{2023~CX1}\\
\hline
Time (UTC) & 2018-06-02 16:44:14.464 \\
Latitude (deg) & $-21.373816 \pm 0.012043$ \\
Longitude (deg) & $24.741922 \pm 0.0.067253$ \\
Speed (km s$^{-1}$) & $16.999 \pm 0.001830$ \\
Flight path angle (deg) & $-25.126 \pm 0.037865$ \\
Heading angle (deg) & $275.195, \pm 0.019960$ \\
1-$\sigma$ semi-major axis (m) & 7090.91 \\
1-$\sigma$ semi-minor axis (m) & 376.29 \\
1-$\sigma$ azimuth (deg) & 100.45 \\
\hline
\end{tabular}
\caption{Impact parameters at 100 km altitude (WGS84 ellipsoid) for asteroid 2018~LA as computed by Aegis. The velocity is given with respect to the ground.}
\label{tab:la_init}
\end{table}

\begin{figure}
    \centering
    \includegraphics[width=1\linewidth]{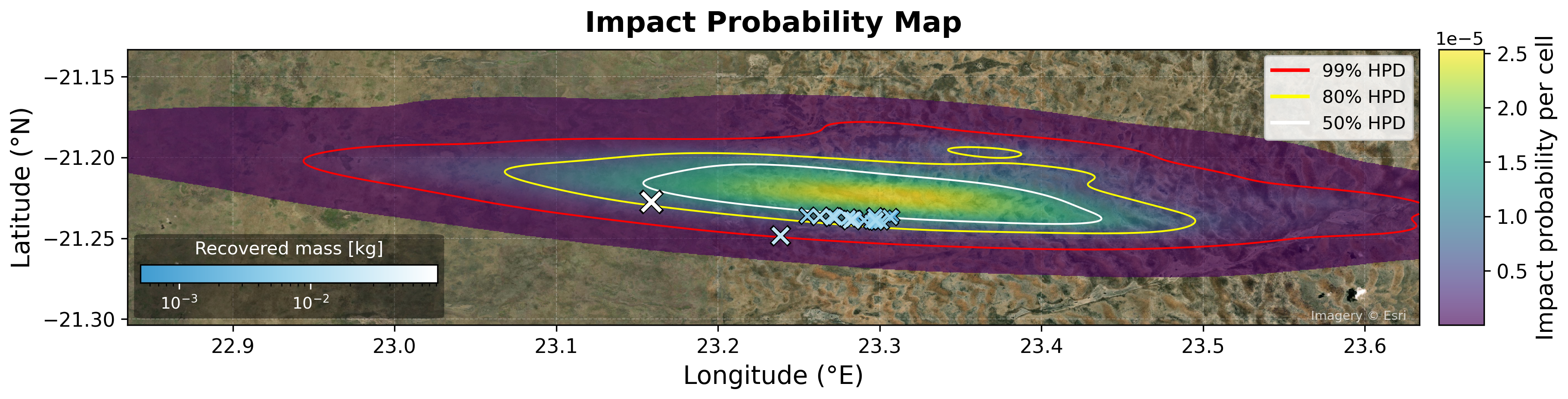}
    \caption{KDE-based mass probability map for 2018~LA. Colors represent the spatial mass probability density, contours denote the highest posterior density regions, and crosses mark recovered meteorites (size and color proportional to mass).}
    \label{fig:2018LA_sf}
\end{figure}

\subsection{Winchcombe fireball}

  This case does not represent an imminent impactor scenario, but we can still use it to demonstrate that the model is applicable to fireball events as well and remains reliable, thus demonstrating its generality.

  The Winchcombe fireball occurred on 28 February 2021 and led to the recovery of approximately 0.6~kg of CM2 carbonaceous chondrite meteorites (the simulations are therefore performed using the carbonaceous
chondrite parameter set; ordinary chondrite parameters are used as default for all other cases) near Winchcombe, in South-West England. The large number of recovered fragments provides a suitable dataset for validation.

  The initial conditions adopted in the simulations are derived from Tables~2 and~3 of the published trajectory reconstruction in \cite{2024M&PS...59..927M}. While the nominal Aegis workflow typically initialises trajectories at the conventional atmospheric-entry altitude of 100~km, in this specific case the simulation is started directly at the first point of the observed trajectory (about 90.6~km), corresponding to the beginning of the luminous fireball phase.
The Monte Carlo simulations produce a strewn field consistent with the observed distribution (see Fig.~\ref{fig:kdewinchcombe}), where all but two fragments (78\%) fell in the 99.9\% HPD. The entry-state uncertainties adopted here from \cite{2024M&PS...59..927M} are very small, resulting in a narrow predicted strewn field that may be slightly optimistic and could explain why some recovered meteorites fall outside the highest-probability region. The meteorite recovery locations were obtained from \citet{2024M&PS...59..973R}.
\begin{figure}
    \centering
\includegraphics[width=1\linewidth]{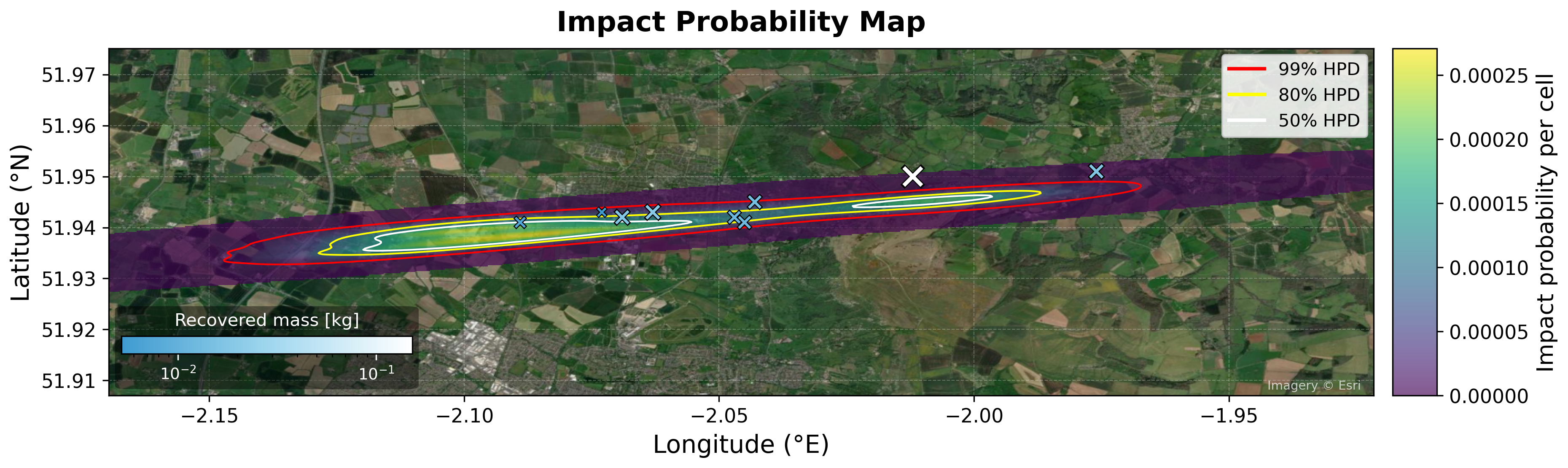}
    \caption{KDE-based impact probability map for the Winchcombe meteoroid. Colors show the spatial impact probability density, contours indicate the highest posterior density regions, and crosses mark recovered meteorites (size and color proportional to mass).}
    \label{fig:kdewinchcombe}
\end{figure}

\subsection{Chelyabinsk asteroid}

  This case does not represent an imminent impactor scenario, as
it was reconstructed from fireball observations. However, it is the only
large-entry event currently available to test the model against a
well-defined strewn field. As previously discussed, the Chelyabinsk
event caused numerous injuries, highlighting the importance of being able to predict in advance where the airburst may occur and where the fragments might fall if a similar object were detected today. For this
reason, we include this case to verify that the model can also be applied
to decametre-asteroid entries while maintaining good reliability.

  The Chelyabinsk event occurred on 15 February 2013, when a small asteroid entered Earth's atmosphere over Russia, producing a bright fireball and airburst without prior detection. The object was estimated to be 15-20 meters in diameter with a pre-atmospheric mass of approximately 11,000 tons, entering at velocities around 18 km s$^{-1}$. The event resulted in numerous fragments recovered near Chebarkul Lake, confirming its composition as ordinary chondrite.

  The initial conditions for the simulations are derived from \cite{2013arXiv1304.2410P}, with the entry parameters in Table 1, and material properties and mass in Table 2, as summarized in Table~\ref{tab:chelyabinsk_parameters}. The ablation parameter and the fragmentation parameters were selected within the model to reproduce the recovered-meteorite distribution.

\begin{table}[h]
\centering
\begin{tabular}{lc}
\hline
\textbf{Parameter} & \textbf{Chelyabinsk} \\
\hline
Latitude (deg) & $54.51$ \\
Longitude (deg) & $64.27$ \\
Altitude (km) & $91.83$ \\
Speed (km s$^{-1}$) & $17.5$ \\
Flight path angle (deg) & $-19.9$ \\
Heading angle (deg) & $282.41$ \\
Diameter (m) & $18$ \\
Density (kg m$^{-3}$) & $3600$ \\
Strength (Pa) & $1.0 \times 10^{7}$ \\
Ablation parameter & $7.5 \times 10^{-8}$ \\
Dust fraction & $0.75$ \\
$\beta$ & $0.4$ \\
Largest-fragment mass fraction & $0.1$ \\
Core-retention fraction & $0.85$ \\
\hline
\end{tabular}
\caption{Input parameters used for the representative Chelyabinsk simulation of Fig.~\ref{fig:nominalchelyabinsk}.}
\label{tab:chelyabinsk_parameters}
\end{table}

  The Monte Carlo simulations yield excellent results, as shown
in Fig.~\ref{fig:kdechelyabinsk}. One of the Monte Carlo runs, for which the full set of initial conditions is reported in Tab.~\ref{tab:chelyabinsk_parameters}, notably reached a median error
of approximately 115~m (Fig.~\ref{fig:nominalchelyabinsk}). 
This simulation predicts a dominant breakup occurring at an 
altitude of about 27~km, producing one large surviving core together 
with numerous smaller fragments. Among these, five of them undergo 
further secondary fragmentation at lower altitude. The resulting mass 
distribution agrees very well with the observed meteorites. In 
particular, the largest fragment is reproduced with high accuracy, 
with a simulated core mass of 602~kg compared to an observed value of 
about 600~kg. The agreement is also good for the smaller fragments, and 
the overall fragmentation behaviour and breakup altitude are consistent 
with the description reported in \cite{popova2013}.
Overall,
about 84.5\% of the recovered meteorites fall within the 95\% probability
contour, while 97\% are contained within the 99\% contour. The largest
recovered fragment, whose mass exceeds that of the smaller meteorites by
roughly two orders of magnitude, is reproduced very accurately in terms
of location but lies outside the 95\% region. This occurs because the KDE
map represents the density of impact points, treating all fragments
equally regardless of mass. If the probability map were instead weighted
by fragment mass, the density would be strongly concentrated near this
largest fragment, and the highest-probability region would shift toward
its location.
 The meteorite recovery locations were obtained from the LPI Strewn Field database \citep{lpi_tc3}.

\begin{figure}
    \centering
\includegraphics[width=1\linewidth]{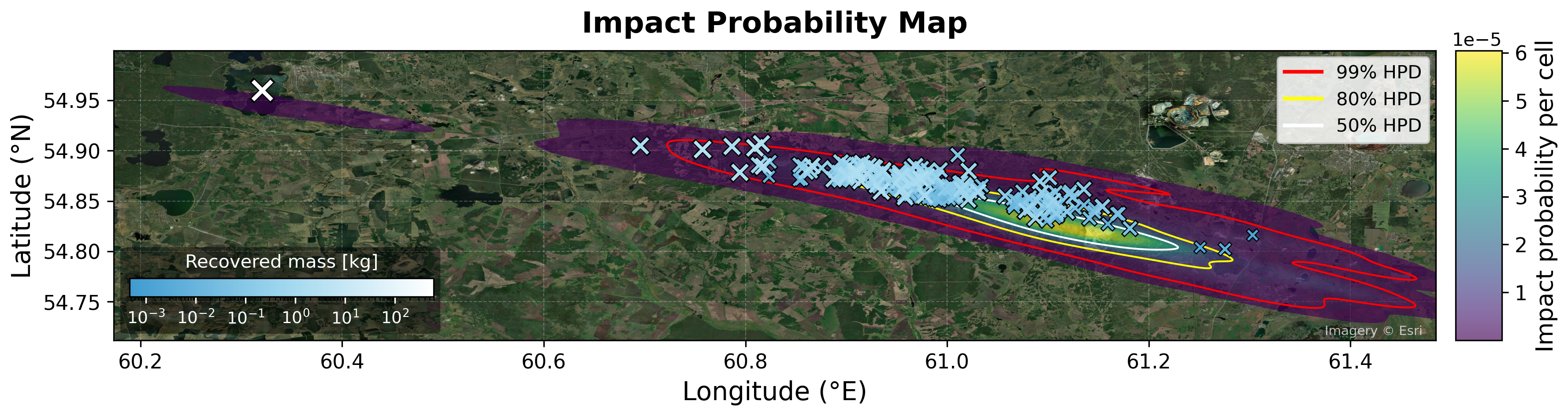}
    \caption{KDE-based impact probability map for the Chelyabinsk asteroid. Colors show the spatial impact probability density, contours indicate the highest posterior density regions, and crosses mark recovered meteorites (size and color proportional to mass).}
    \label{fig:kdechelyabinsk}
\end{figure}
\begin{figure}
    \centering
\includegraphics[width=0.7\linewidth]{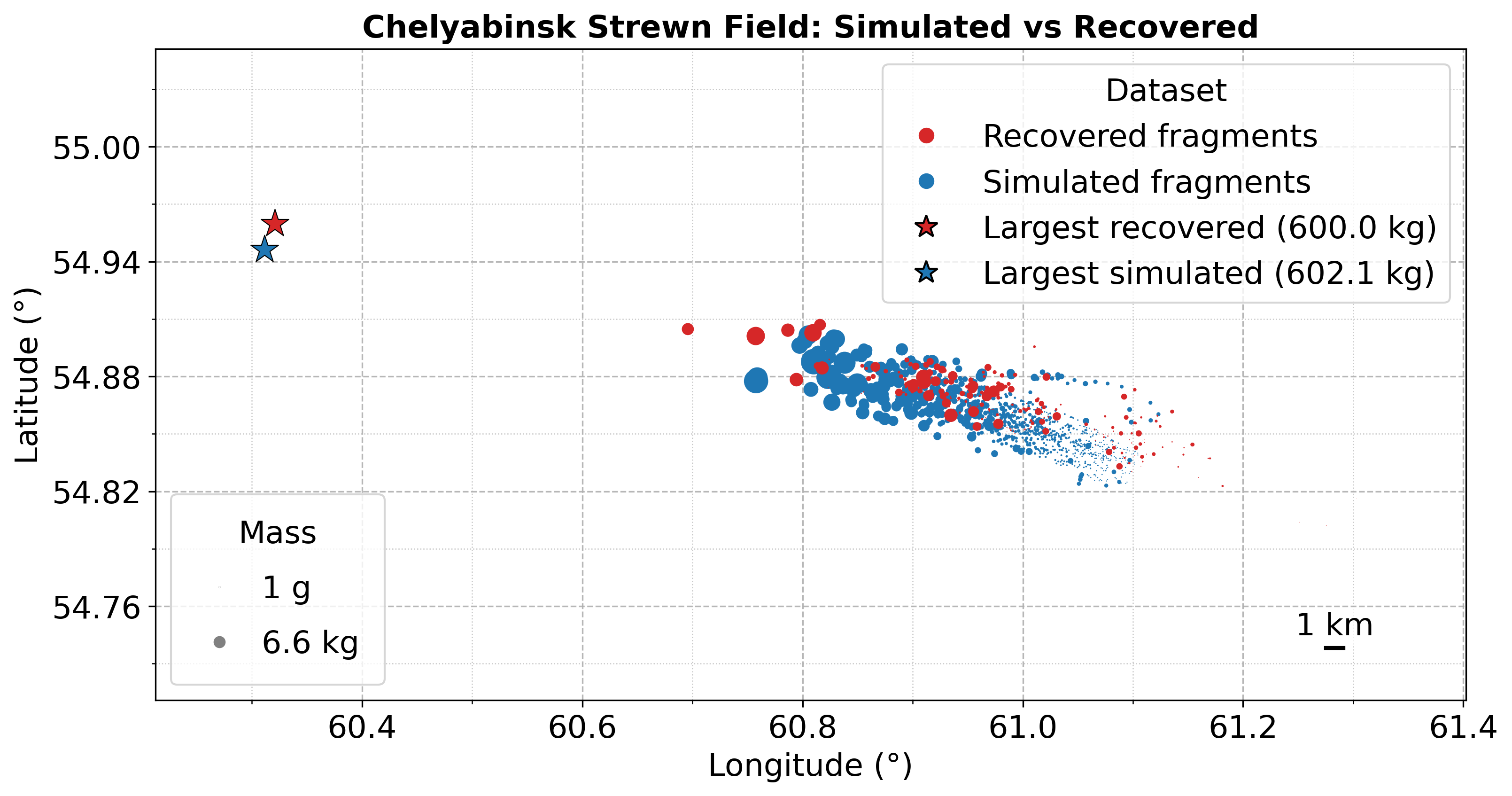}
\caption{Monte Carlo realisation providing the closest match to the recovered meteorites for the Chelyabinsk fall. Simulated fragment impacts are shown together with recovered meteorite locations; marker size is proportional to fragment mass.}
\label{fig:nominalchelyabinsk}
\end{figure}

\section{Discussion}
\label{s:discussion}
  The results obtained in this work show that the model can reproduce the geometry, extent, and orientation
of observed strewn fields across different entry conditions. Compared to the closely related \emph{ab initio} approach of \citet{2025Icar..42516345C}, the present model introduces several key improvements. First, the atmospheric entry is treated within a full Monte Carlo framework rather than through nominal simulations with fixed parameters, allowing the uncertainty in orbital state, physical and atmospheric properties to be propagated consistently to the ground. Second, fragmentation is modeled as a statistical cascade that conserves mass and generates a continuous distribution of fragment sizes, instead of a single discrete breakup with a small set of predefined representative masses. The model also includes lateral ejection velocities at fragmentation, three-dimensional wind fields, digital elevation models, and the possibility of multiple fragmentation episodes. These elements allow the method to produce a probabilistic strewn field distribution rather than a single deterministic solution, improving the realism of the result and extending applicability to larger events such as Chelyabinsk-scale entries.

  The Monte Carlo ensemble highlights how
multiple parameters jointly shape the strewn field, with material
strength $S$ and atmospheric winds playing especially important roles.
Even small variations in $S$ can shift the breakup altitude by several
kilometres, thereby modifying the along-track dispersion and altering
the time fragments remain exposed to atmospheric winds. The latter,
particularly horizontal shear in the troposphere and lower stratosphere,
influence the dark-flight phase, where they can induce cross-track
displacements comparable to or even exceeding the lateral spread
generated directly during fragmentation. This interplay demonstrates
that the final strewn field pattern is governed by combinations of
parameters rather than by a strict hierarchy of individual effects.
Fragment mass modulates the sensitivity to both breakup dynamics and
wind forcing: larger bodies, having smaller area-to-mass ratios, are
less affected by drag and lateral ejection, resulting in smaller
positional uncertainties. In contrast, smaller fragments experience
stronger aerodynamic deceleration and retain a larger fraction of their
transverse breakup velocity, producing broader and less predictable
distributions. Entry trajectory inclination further influences the outcome: steep entries 
shorten the atmospheric path and cause earlier deceleration to terminal 
velocity, reducing the duration of dark flight and thus limiting wind-driven 
lateral transport; conversely, shallow entries extend the atmospheric path, 
delay the onset of dark flight, amplify cumulative wind effects, and enlarge 
the search area.

  While an overall good agreement with observed strewn fields is
achieved, residual discrepancies remain and warrant consideration. These
can be separated into epistemic and aleatory sources of uncertainty. The
former arise from structural simplifications of the model. Fragments are
assumed spherical, so lift and side forces are neglected, and rotational
dynamics are not modeled; irregular shapes and tumbling could introduce
small but systematic lateral deviations. The ablation coefficient is
treated as constant during the luminous phase, whereas laboratory and
observational studies suggest that heat-transfer efficiency may vary
with altitude and flow regime. Fragmentation is described through a
statistical cascade with strength scaling, but no explicit
aerothermodynamic--structural coupling is included. Aerodynamic
interactions between fragments are also neglected: wake confinement,
shielding, and drag coupling within fragment clouds are not modeled.
Similarly, real meteoroids likely possess more heterogeneous internal
structures than represented by the adopted Weibull-type scaling law;
rubble-pile textures, macroscopic voids, or preferential fracture
planes could alter the fragmentation sequence.

  Despite these simplifications, their practical impact on the
predicted ground distribution appears limited for the cases studied
here. The model reproduces the main observable features of the strewn
fields even without resolving high-fidelity fluid--structure dynamics.
Future developments could mitigate these limitations by introducing
shape-dependent drag and lift coefficients, simplified rotational
dynamics, altitude-dependent ablation efficiencies, or parameterized
wake-interaction corrections.

  Aleatory uncertainties, in contrast, reflect the intrinsic
variability of the physical environment and input parameters. The
atmospheric model is based on GFS data with a horizontal resolution of
approximately 0.25$^\circ$, corresponding to grid spacings of the order
of 25~km. While adequate for capturing large-scale wind patterns, this
resolution is coarse relative to the spatial scale of typical strewn
fields. Although ad-hoc post-impact atmospheric reconstructions can be
computed for specific events \citep[see e.g.][]{2021MNRAS.501.1215G},
GFS data represent the best operationally available option. Small-scale
turbulence, convective cells, and localized wind structures are not
resolved, and their cumulative effect during dark flight may account for
part of the residual dispersion. The stochastic sampling of strength,
density, albedo, and fragmentation parameters further introduces
variability that cannot be deterministically reduced in the absence of
additional observational constraints, which are often lacking for
imminent impactors. In particular, the pre-impact strength $S$ remains
the least constrained parameter. Independent estimates derived from
light-curve behaviour, rotational state, taxonomy, or spectral class
would narrow the admissible range of breakup altitudes and improve
predictions, but these are typically available only after impact, if at
all.

  The ``best'' individual realisation within the Monte Carlo
ensemble should therefore not be interpreted as a unique reconstruction
of the event. When no additional observational constraints are
available, different combinations of strength, fragmentation parameters,
and atmospheric conditions can lead to very similar strewn field
patterns. In this sense, the inverse problem is degenerate: different
fragmentation histories may produce nearly identical ground
distributions, making it impossible to identify a single physical
scenario. The situation improves when fireball observations are
available.

  Importantly, this lack of uniqueness does not prevent the
model from being useful in practice. Even though breakup altitude may
vary significantly across the Monte Carlo runs, the predicted impact
area remains relatively compact. Different fragmentation scenarios tend
to project onto similar ground patterns once atmospheric drag and winds
are taken into account. As a result, the probability map is stable
enough to define a well-delimited search region. While the model does not reconstruct every detail of the fragmentation process, it provides impact-area estimates and fragment-property distributions that are robust enough to support rapid-response analysis, meteorite recovery operations, and preliminary civil protection assessment. The present framework provides several quantities that are directly relevant to impact response assessment, including the spatial probability distribution of surviving fragments, their terminal masses, and their impact velocities. These outputs can be used to identify areas where ground-reaching material is more likely and to distinguish low-mass, low-energy fragments from rarer but potentially more hazardous surviving masses. However, the conversion of these physical predictions into operational civil protection actions is out of the scope of this work. Decisions such as evacuation, sheltering, road closure, or protection of sensitive infrastructure require an additional exposure and vulnerability layer, including population distribution, land use, infrastructure maps, acceptable risk thresholds, and the practical cost of intervention. Therefore, the model should be regarded as a pre-impact situational awareness and meteorite recovery tool that can inform planetary defence response for small imminent impactors, rather than as a stand-alone evacuation decision system.

\section{Conclusions}
\label{s:conclusions}

  In recent years, the atmospheric entry of meteoroids detected only a few hours before impact has become a recurrent occurrence, reflecting the increasing sensitivity and cadence of modern survey systems \citep{2025Icar..42516345C}. 
Here we presented a fully numerical Monte Carlo model for predicting meteorite strewn fields directly from pre-impact orbital solutions, which uses a-priori constraints to model physical uncertainties. 
The fall model couples luminous flight, cascading fragmentation, and dark flight within a unified dynamical scheme. Unlike traditional approaches, it does not require fireball triangulation or post-event calibration, enabling the computation of strewn field probability maps prior to atmospheric entry. The model was integrated into the ESA automated pipeline for imminent impactors, enabling the automated prediction of strewn fields of meteoroids when they are still on the NEOCP.

  The model was validated against multiple well-documented falls. Here we reported the results obtained for 2023~CX1 and 2008~TC3. Although they are not imminent impactors, we further tested the model on the Winchcombe fireball, and on the Chelyabinsk event. These tests demonstrate that the framework reliably reproduces the orientation, spatial extent, and mass-dependent dispersion of observed strewn fields. In favourable cases, nominal solutions reach median ground-distance discrepancies on the order of 100–200 m. These distances correspond to areas that are directly searchable on foot by coordinated teams, or even surveyable using drones \citep{2022ApJ...930L..25A}.

  Although the framework was designed for imminent impactors detected through telescopic surveys, it is not restricted to initialization from heliocentric orbital solutions. The numerical propagation can be started at any altitude and from arbitrary initial conditions in position, velocity, and mass. As shown in the validation cases, the model performs consistently when initialized from reconstructed fireball states and remains applicable to larger objects such as Chelyabinsk-scale asteroids. In this sense, the software constitutes a general atmospheric entry and strewn field propagation tool, adaptable across a wide range of object sizes and observational scenarios, provided suitable initial conditions are supplied.

  The integration of the model within the ESA Meerkat/Aegis pipeline demonstrates that routine, automated strewn field prediction for imminent impactors is technically feasible. The ability to compute probability maps hours ahead of impact opens the possibility for better coordination with local emergency response authorities, pre-positioning teams for fireball observations, and planning rapid recovery campaigns. For fragile carbonaceous meteorites, early recovery reduces terrestrial alteration and contamination, preserving scientific value. For larger impactors, timely localisation of potential impact corridors may also support civil protection assessments. The approach is not intended to completely replace fireball observation networks, rather to complement them. It remains effective in daylight events, sparsely instrumented regions, or cases lacking atmospheric observations, while still benefiting from additional constraints when fireball data are available.

\clearpage
\printcredits

\section*{Acknowledgements}
This research has made use of data and/or services provided by the International Astronomical Union's Minor Planet Center. We thank the referees for their valuable comments, which helped improving the quality of the paper.

\bibliographystyle{cas-model2-names}

\bibliography{holyBib}

\end{document}